\documentclass{article}
\usepackage{epsfig}
\usepackage[hang,nooneline]{subfigure}
\date{\today}

\usepackage{amsmath}
\usepackage{amssymb}

\begin{document}
\title{Hairy charged Gauss-Bonnet solitons and black holes}
\author{{\large Yves Brihaye \footnote{email: yves.brihaye@umons.ac.be} }$^{\ddagger}$ and 
{\large Betti Hartmann \footnote{email: b.hartmann@jacobs-university.de}}$^{\dagger}$
\\ \\
$^{\ddagger}${\small Physique-Math\'ematique, Universite de
Mons-Hainaut, 7000 Mons, Belgium}\\ 
$^{\dagger}${\small School of Engineering and Science, Jacobs University Bremen, 28759 Bremen, Germany}  }

\date{}
\newcommand{\dd}{\mbox{d}}
\newcommand{\tr}{\mbox{tr}}
\newcommand{\la}{\lambda}
\newcommand{\ka}{\kappa}
\newcommand{\f}{\phi}
\newcommand{\vf}{\varphi}
\newcommand{\F}{\Phi}
\newcommand{\al}{\alpha}
\newcommand{\ga}{\gamma}
\newcommand{\de}{\delta}
\newcommand{\si}{\sigma}
\newcommand{\bomega}{\mbox{\boldmath $\omega$}}
\newcommand{\bsi}{\mbox{\boldmath $\sigma$}}
\newcommand{\bchi}{\mbox{\boldmath $\chi$}}
\newcommand{\bal}{\mbox{\boldmath $\alpha$}}
\newcommand{\bpsi}{\mbox{\boldmath $\psi$}}
\newcommand{\brho}{\mbox{\boldmath $\varrho$}}
\newcommand{\beps}{\mbox{\boldmath $\varepsilon$}}
\newcommand{\bxi}{\mbox{\boldmath $\xi$}}
\newcommand{\bbeta}{\mbox{\boldmath $\beta$}}
\newcommand{\ee}{\end{equation}}
\newcommand{\eea}{\end{eqnarray}}
\newcommand{\be}{\begin{equation}}
\newcommand{\bea}{\begin{eqnarray}}

\newcommand{\ii}{\mbox{i}}
\newcommand{\e}{\mbox{e}}
\newcommand{\pa}{\partial}
\newcommand{\Om}{\Omega}
\newcommand{\vep}{\varepsilon}
\newcommand{\bfph}{{\bf \phi}}
\newcommand{\lm}{\lambda}
\def\theequation{\arabic{equation}}
\renewcommand{\thefootnote}{\fnsymbol{footnote}}
\newcommand{\re}[1]{(\ref{#1})}
\newcommand{\R}{{\rm I \hspace{-0.52ex} R}}
\newcommand{\N}{{\sf N\hspace*{-1.0ex}\rule{0.15ex}%
{1.3ex}\hspace*{1.0ex}}}
\newcommand{\Q}{{\sf Q\hspace*{-1.1ex}\rule{0.15ex}%
{1.5ex}\hspace*{1.1ex}}}
\newcommand{\C}{{\sf C\hspace*{-0.9ex}\rule{0.15ex}%
{1.3ex}\hspace*{0.9ex}}}
\newcommand{\eins}{1\hspace{-0.56ex}{\rm I}}
\renewcommand{\thefootnote}{\arabic{footnote}}

\maketitle

\bigskip

\begin{abstract}
We study the stability of $(4+1)$-dimensional charged Gauss-Bonnet black holes and solitons.
We observe an instability related to the condensation of a scalar field
and construct explicit ``hairy'' black hole and soliton solutions of the full system of coupled
field equations. 
We investigate the cases of a massless scalar field as well as that of
a tachyonic scalar field. 
The solitons with scalar hair exist for a particular range of the charge and the gauge coupling. This range is such that for intermediate values of the gauge coupling
a ``forbidden band'' of charges for the hairy solitons exists. 
We also discuss the behaviour of the black holes with scalar hair 
when changing the horizon radius
and/or the gauge coupling and find that various 
scenarios at the approach of a limiting solution appear. One observation is that
hairy Gauss-Bonnet
black holes {\it never} tend to a regular soliton solution in the limit of vanishing horizon 
radius. We also prove that extremal Gauss-Bonnet black holes can {\it not} carry massless
or tachyonic scalar hair and show that our solutions tend to their planar counterparts for large charges.
 
\end{abstract}
\section{Introduction}
The gravity--gauge theory duality \cite{ggdual} has attracted a lot of attention
in the past years. The most famous
example is the AdS/CFT correspondence \cite{adscft} which states that a gravity
theory in a $d$-dimensional
Anti-de Sitter (AdS) space--time is equivalent to a Conformal Field Theory (CFT)
on the $(d-1)$-dimensional boundary of AdS.
Recently, this correspondence has been used to describe so-called holographic
superconductors with the help of black holes in
higher dimensional AdS space--time \cite{gubser,hhh,horowitz_roberts,reviews}.
In most cases $(3+1)$-dimensional
black holes with planar horizons ($k=0$) were
chosen to account for the fact that high temperature superconductivity is mainly
associated to 2-dimensional layers within the material. 
The basic idea is that at low temperatures a planar black hole in asymptotically
AdS becomes unstable to the condensation
of a charged scalar field. The hairy black hole is the gravity dual of the
superconductor. The main point here is that
this instability occurs due to the fact that the scalar field is charged and its
effective mass
drops below the Breitenlohner-Freedman (BF) bound \cite{bf} for sufficiently low
temperature of the black hole hence
spontaneously breaking the U(1) symmetry.
Surprisingly, however, the scalar condensation can also occur for uncharged
scalar fields in the $d$-dimensional planar Reissner-Nordstr\"om-AdS (RNAdS)
black hole 
space-time \cite{hhh}. This is a new type of instability that is not connected
to a spontaneous symmetry breaking as in the charged case.
Rather it is related to the fact that the planar RNAdS black hole possesses an
extremal limit
with vanishing Hawking temperature and near-horizon geometry 
AdS$_2\times {\mathbb{R}}^{d-2}$ (with $d \ge 4$)
\cite{Robinson:1959ev,Bertotti:1959pf,Bardeen:1999px}. For scalar field masses 
larger than the $d$-dimensional BF bound, but smaller than the $2$-dimensional
BF bound the near-horizon geometry
becomes unstable to the formation of scalar hair, while the asymptotic AdS$_d$
remains stable \cite{hhh}.
The fact that the near-horizon geometry of extremal black holes is a topological
product of two manifolds with constant
curvature has led to the development of the entropy function formalism
\cite{Sen:2005wa,sen2,dias_silva}. 
In \cite{Dias:2010ma} the question of the condensation of an uncharged scalar
field on uncharged black holes in $(4+1)$ dimensions has been addressed.
As a toy model for the rotating case, static black holes with hyperbolic
horizons ($k=-1$) were discussed.
In contrast to the uncharged, static black holes with flat ($k=0$) or spherical
($k=1$) horizon topology
hyperbolic black holes possess an extremal limit with near-horizon geometry
AdS$_2\times H^3$ and it has been shown
by numerical construction that the black holes form scalar hair close to
extremality. 

Higher curvature corrections appear naturally in the low energy effective action
of string theory \cite{zwiebach}. In more than
four dimensions the quadratic correction is often chosen to be the Gauss-Bonnet
(GB) term, which has the property
that the equations of motion are still second order in derivatives of the metric
functions. As such explicit
solutions of the equations of motion are known. The first example of static,
spherically symmetric and asymptotically flat
black hole solutions in GB gravity were given for the uncharged case in
\cite{deser,Wheeler:1985nh} and for the charged case in 
\cite{wiltshire}.
Moreover, the corresponding solutions in asymptotically Anti-de Sitter (AdS)
\cite{Cai:2001dz,Cvetic:2001bk,Cho:2002hq,Neupane:2002bf}
as well as de Sitter (dS) space-times \cite{Cai:2003gr} have been studied. In
most cases, black holes not only with
spherical ($k=1$), but also with flat ($k=0$) and hyperbolic ($k=-1$) horizon
topology have been considered.
Moreover, the thermodynamics of these black holes has been studied in detail
\cite{Cho:2002hq,Neupane:2002bf,Neupane:2003vz}
and the question of negative entropy for certain GB black holes in dS and AdS
has been discussed \cite{Cvetic:2001bk,Clunan:2004tb}.

Uncharged $d$-dimensional Gauss Bonnet black holes with hyperbolic horizon topology possess also a
regular extremal limit such that the near horizon geometry contains an AdS$_2$ factor.
For scalar fields with masses above the $d$-dimensional BF bound, but below the 2-dimensional
BF bound one would expect the near-horizon geometry to become unstable to scalar hair formation. This was demonstrated
in \cite{hartmann_brihaye3}.
In particular, it was shown 
that the radius of the near-horizon AdS$_2$ decreases with
increasing Gauss-Bonnet coupling and tends to zero in the Chern-Simons limit \cite{hartmann_brihaye3}.

Interestingly, there seems to be
a contradiction between the holographic superconductor approach and the
Coleman-Mermin-Wagner theorem \cite{CMW} 
which forbids spontaneous symmetry
breaking in $(2+1)$ dimensions at finite temperature. Consequently, it has been
suggested that 
higher curvature corrections and in particular GB terms should 
be included on the gravity side and holographic GB superconductors in $(3+1)$
dimensions have been
studied \cite{Gregory:2009fj}. However, though the critical temperature gets
lowered
when including GB terms, condensation cannot be suppressed -- not even when
including backreaction \cite{Brihaye:2010mr,Barclay:2010up,Siani:2010uw}.

In \cite{dias2}
static, spherically symmetric black hole and soliton solutions to Einstein-Maxwell 
theory coupled to a charged, massless scalar field 
in (4+1)-dimensional AdS space-time have been studied. The existence
of solitons in global AdS was discovered in \cite{basu}, 
where a perturbative approach was taken. In \cite{dias2} it was shown that
solitons can have arbitrarily large charge for large enough gauge coupling, 
while for small gauge coupling the solutions exhibit a spiraling behaviour towards
a critical solution with finite charge and mass. The stability of RNAdS
solutions was also studied in this paper. It was found that for small gauge coupling RNAdS black holes
are never unstable to condensation of a massless, charged scalar field, while for 
intermediate gauge couplings RNAdS black holes become unstable for sufficiently large charge.
For large gauge coupling RNAdS black holes are unstable to formation of massless scalar hair for
all values of the charge. Moreover, it was observed that for large gauge coupling and
small charges the solutions exist all the way down to vanishing horizon. The limiting
solutions are the soliton solutions mentioned above. On the other hand for large charge the
limiting solution is a singular solution with vanishing temperature and finite entropy, which
is not a regular extremal black hole \cite{fiol1}. These results were extended to a tachyonic scalar 
field as well as to the rotating case \cite{brihaye_hartmannNEW}.
Recently, solutions in asymptotically global AdS in 4 dimensions have been studied in \cite{menagerie}.
It was pointed out  that the solutions tend to their planar counterparts for large charges since 
in that case the solutions can become
comparable in size to the AdS radius.

In this paper, we are interested in the condensation of a charged tachyonic or massless scalar 
fields on charged static black holes and solitons in (4+1)-dimensional Anti-de Sitter space-time. 
We reinvestigate the case of Einstein gravity and point out additional features.
Then we extend our results to include Gauss-Bonnet corrections.

Our paper is organized as follows: we present the model in Section 2 and discuss the planar limit of
our solutions in Section 3. We then present soliton solutions with scalar hair 
in Section 4 and ``hairy'' black holes in Section 5.
We conclude and summarize in Section 6.

\section{The model}

In this paper, we are studying the formation of scalar hair on electrically charged black holes 
and solitons in $(4+1)$-dimensional Anti--de Sitter space--time. 
The action reads~:
\begin{equation}
S= \frac{1}{16\pi G} \int d^5 x \sqrt{-g} \left(R -2\Lambda + 
\frac{\alpha}{2}\left(R^{MNKL} R_{MNKL} - 4 R^{MN} R_{MN} + R^2\right) + 16\pi G {\cal L}_{\rm matter}\right) \ ,
\end{equation}
where $\Lambda=-6/L^2$ is the cosmological constant, $\alpha$ the Gauss--Bonnet coupling
and $M,N,K,L=0,1,2,3,4$. 
${\cal L}_{\rm matter}$ denotes the matter Lagrangian~:
\begin{equation}
{\cal L}_{\rm matter}= -\frac{1}{4} F_{MN} F^{MN} - 
\left(D_M\psi\right)^* D^M \psi - m^2 \psi^*\psi  \ \ , \ \  M,N=0,1,2,3,4  \ ,
\end{equation}
where $F_{MN} =\partial_M A_N - \partial_N A_M$ is the field strength tensor and
$D_M\psi=\partial_M \psi - ie A_M \psi$ is the covariant derivative.
$e$ and $m^2$ denote the electric charge and mass of the scalar field $\psi$, respectively.

The coupled gravity and matter field equations are obtained from the variation of the
action with respect to the matter and metric fields, respectively, and read
\begin{equation}
 G_{MN} + \Lambda g_{MN} + \frac{\alpha}{2} H_{MN}=8\pi G T_{MN} \ ,  \ M,N=0,1,2,3,4 \ ,
\end{equation}
where $H_{MN}$ is given by
\begin{equation}
 H_{MN}= 2\left(R_{MABC}R_N^{ABC} - 2 R_{MANB}R^{AB} - 2 R_{MA}R^{A}_N + R R_{MN}\right)
- \frac{1}{2} g_{MN} \left(R^2 - 4 R_{AB}R^{AB} + R_{ABCD} R^{ABCD}\right)
\end{equation}
and $T_{MN}$ is the energy-momentum tensor
\begin{equation}
 T_{MN}=g_{MN} {\cal L}_{\rm matter} - 2\frac{\partial {\cal L}_{\rm matter}}{\partial g^{MN}} \ .
\end{equation}

We choose the following Ansatz for the metric~:
\begin{equation}
ds^2 = - f(r) a^2(r) dt^2 + \frac{1}{f(r)} dr^2 + \frac{r^2}{L^2} d\Omega^2_{3}
\end{equation}
where $f$ and $a$ are functions of $r$ only.
For the electromagnetic field and the scalar field we have \cite{hhh}~:
\begin{equation}
A_{M}dx^M = \phi(r) dt \  \  \  , \   \   \   \psi=\psi(r)
\end{equation}
such that the solutions possess only electric charge.

The equations of motion then read~:
\begin{eqnarray}
\label{eq1}
     f' &=& 2r \frac{1-f+ 2 r^2/{L^2}}{r^2 + 2 \alpha(1-f)} 
     - \gamma \frac{r^3}{2 f a^2} 
     \left(\frac{2 e^2 \phi^2 \psi^2 + f (2 m^2 a^2 \psi^2 + \phi'^2) + 
2 f^2 a^2 \psi'^2}{r^2 + 2 \alpha (1-f))}\right) \\
\label{eq2}
        a' &=& \gamma \frac{r^3(e^2 \phi^2 \psi^2 + a^2 f^2 \psi'^2)}{a f^2(r^2+ 2 \alpha(1-f))}\\
\label{eq3}
   \phi'' &=& - \left(\frac{3}{r} - \frac{a'}{a}\right) \phi' +
2 \frac{e^2 \psi^2}{f} \phi \\
\label{eq4}
    \psi'' &=& -\left(\frac{3}{r} + \frac{f'}{f} + \frac{a'}{a}\right) \psi'- 
\left(\frac{e^2 \phi^2}{f^2 a^2} - \frac{m^2}{f}\right) \psi
\end{eqnarray}
where $\gamma=16\pi G$. Here and in the following the prime denotes
the derivative with respect to $r$. 
These equations
depend on the following independent constants: Newton's constant $G$, 
the cosmological constant $\Lambda$ (or Anti-de Sitter radius $L$) and the charge $e$ and mass $m$ of the 
scalar field as well as on the Gauss-Bonnet coupling $\alpha$. 
The system possesses two scaling symmetries:
\begin{equation}
\label{scaling1}
 r\rightarrow \lambda r \ \ \ , \ \ \ t\rightarrow \lambda t \ \ \ , \ \ \ L\rightarrow \lambda L
\ \ \ , \ \ \ e\rightarrow e/\lambda \ \ \ , \ \ \ \alpha\rightarrow \lambda^2 \alpha
\end{equation}
as well as 
\begin{equation}
\label{scaling2}
 \phi \rightarrow \lambda \phi \ \ \ , \ \ \ \psi \rightarrow \lambda \psi \ \ \ , \ \ \ 
e\rightarrow e/\lambda \ \ \ ,  \ \ \ \gamma\rightarrow \gamma/\lambda^2
\end{equation}
which we can use to set $L=1$ and $\gamma$ to some fixed value without loosing generality.

Asymptotically, we want the space--time to be that of global AdS, i.e. we can choose
$a(r\rightarrow\infty)\rightarrow 1$. Other choices of the asymptotic value
of $a(r)$ would simply correspond to a rescaling of the time coordinate. The matter fields on the other hand obey~:
\begin{equation} 
  \phi(r\gg 1) = \mu - Q/r^2  \ \ , \ \ 
  \psi(r\gg 1) = \frac{\psi_{-}}{r^{\lambda_{-}}} + \frac{\psi_{+}}{r^{\lambda_{+}}} \ \
\label{decay}
\end{equation}
with
\begin{equation}
\label{lambda}
       \lambda_{-} = 2 - \sqrt{4 +m^2 L_{\rm eff}^2} \ \ , \ \ \lambda_{+} = 
2 + \sqrt{4 + m^2 L_{\rm eff}^2} \ \ , \ \ 
       L_{\rm eff}^2 \equiv \frac{2 \alpha}{1 - \sqrt{1 - 4 \alpha/L^2}} 
       \sim L^2 \left(1  -  \alpha/ L^2 + O(\alpha^2)\right)   \ .
\end{equation}
Note that the value of the Gauss--Bonnet coupling $\alpha$ 
is bounded from above~: $\alpha \leq L^2/4$ where $\alpha=L^2/4$ is the Chern-Simons limit. In the following we 
will be interested in two cases: that of a tachyonic scalar field with $m^2=-3/L^2$ and
that of a massless scalar field with $m^2=0$. Note that while for the tachyonic case, the
fall-off of the scalar field will depend on $\alpha$, it is independent of $\alpha$ in the
massless case. 

The parameters $\mu$, $Q$ are the chemical potential and the electric charge, respectively. 
In the following we will choose $\psi_{-}=0$. 
$\psi_{+}$ will correspond to the expectation value $\langle{\cal O}\rangle$ of the operator
${\cal O}$ which in the context of the gauge theory--gravity duality is dual
to the scalar field on the conformal boundary of AdS.
Note that in comparison to the case of holographic superconductors (see e.g. \cite{hhh}
and reference therein)
the space-time in this study possesses spherically symmetric sections for constant $r$ and $t$.
In other words, for $r\rightarrow \infty$ our space-time corresponds to global
AdS. 

The energy of the solution is given is terms of the coefficients that appear in the
asymptotic fall-off of the metric functions which reads
\begin{equation}
  f(r\gg 1) =  1+ \frac{r^2}{L_{eff}^2} + \frac{f_2}{r^2} + \dots \ \ , \ \ 
 a(r\gg 1) = 1 + \frac{c_4}{r^4} + \dots \ .
\end{equation}
$f_2$ and $c_4$ are constants that have to be determined numerically and that
depend on the couplings in the model.
The energy $E$ then reads
\begin{equation}
   E = \frac{V_3}{8 \pi G} 3 M \ \ \ , \ \ \ M = - \frac{f_2}{2} \sqrt{1 - 4 \alpha/ L^2} \ .
\end{equation}

No explicit solutions to the system of equations (\ref{eq1})-(\ref{eq4}) are known
for $\psi(r) \neq 0$. The solutions have to be constructed numerically. 
We have done this by employing a collocation method for boundary-value ordinary
differential equations, equipped with an adaptive mesh selection procedure \cite{colsys}.
In the following, we will use the rescalings (\ref{scaling1}), (\ref{scaling2}) 
and choose $L=1$ and $\gamma$ to some fixed value. 
Note in particular that the scaling (\ref{scaling2}) leads to $e^2 \rightarrow e^2/\gamma$. 
Due to convenience, we will choose $\gamma=\frac{9}{40}$ in the following. Hence, our
$e^2$ will be related to the $e^2$ of \cite{dias2} -- which we denote $\tilde{e}^2$ in the
following -- by $e^2=\frac{9}{40} \tilde{e}^2$.

\section{The planar limit}

In \cite{menagerie} it was pointed out for the $\alpha=0$ limit and in 4-dimensional global AdS
that the solutions can be connected to the planar counterparts in the limit of large charge.
This is possible since the electromagnetic repulsion balances the gravitational attraction present
in AdS space-time. As such the solutions can become comparable in size to the AdS radius $L$.
On the AdS boundary these solutions look similar to the corresponding planar counterparts. We find that
this is also true in 5-dimensional global AdS and including Gauss-Bonnet corrections and we will present
our numerical results for soliton solutions in the follow section. Here, let us demonstrate the idea
using an analytic and electrically charged black hole solution to the equations of motion for 
$\psi(r)\equiv 0$. This reads
\cite{deser,Cai:2001dz,Cvetic:2001bk,Cho:2002hq,Neupane:2002bf}
\begin{equation}
      f(r) = 1 + \frac{r^2}{2\alpha} \left(1-\sqrt{1-\frac{4\alpha}{L^2} + 
\frac{4\alpha M}{r^4} - \frac{4\alpha\gamma Q^2}{r^6}}\right)  \ \ , \ \ a(r)=1 \ \ , \ \ 
      \phi(r) = \frac{Q}{r_h^2} - \frac{Q}{r^2}  \ ,
\label{rn}
\end{equation}
where $M$ and $Q$ are arbitrary integrations constants that can be interpreted as the mass and
the charge of the solution, respectively. In the limit $\alpha\rightarrow 0$, 
the metric function $f(r)$ becomes $f(r)=1 +\frac{r^2}{L^2}-\frac{M}{r^2}+\frac{\gamma Q^2}{r^4}$
and the corresponding solutions are Reissner-Nordstr\"om-Anti-de Sitter (RNAdS) black holes.
First note that for $M=Q=0$ this solution corresponds to global 5-dimensional AdS with AdS radius $L_{\rm eff}$. 
Now let us rewrite the metric function $f(r)$ as follows
\begin{equation}
     f(r) = r^2\left(\frac{1}{r^2} + \frac{1}{2\alpha} \left(1-\sqrt{1-\frac{4\alpha}{L^2} + 
\frac{4\alpha M}{r^4} - \frac{4\alpha\gamma Q^2}{r^6}}\right) \right)
\label{rn2}
\end{equation}
and adapt a rescaling to the conformal AdS patch that reads
\begin{equation}
 r\rightarrow \lambda r \ \ , \ \ t\rightarrow \lambda^{-1} t \ \ , \ \ M \rightarrow \lambda^4 M \ \ , \ \ 
Q\rightarrow \lambda^3 Q \ \ , 
\end{equation}
where $\lambda$ is some arbitrary scaling parameter. 
For $\lambda \rightarrow \infty$ this tends to the planar solution with 
$\lambda^2 d\Omega^2_{3} \rightarrow dx^2 + dy^2 + dz^2$. This corresponds to $M$ and $Q$ going to infinity. In addition
we have the following rescalings
\begin{equation}
 \mu\rightarrow \lambda \mu \ \ \ , \ \ \ \psi_{\pm} \rightarrow \lambda^{\lambda_{\pm}} \psi_{\pm}   \ .
\end{equation}
In other words, whenever we have a solution in global AdS (black hole or soliton) for which any of the parameters
above tends to infinity we approach the planar limit. 
\section{Solitons}
To find soliton solutions of the equations of motion, we have to fix appropriate boundary
conditions at the origin $r=0$. These read
\begin{equation}
f(0) = 1 \ , \ \phi'(0) = 0 \ , \  \psi'(0) = 0 \ ,  \  a'(0)=0 \ 
\end{equation}
and ensure that the metric
and matter functions are regular at the origin.
The solitons can then be 
characterized by the values of the matter and metric functions at the origin
$\phi(0)$, $\psi(0)$, $a(0)$ which depend on the choice of $e^2$, $Q$, $\alpha$ and $m^2$.

For $\alpha=0$ the existence
of solitons in global AdS was discovered in \cite{basu}, 
where a perturbative approach was taken. In \cite{dias2} it was shown that
solitons can have arbitrarily large charge $Q$ for large enough gauge coupling $e^2$,
while for small gauge coupling the solutions exhibit a spiraling behaviour towards
a critical solution with finite charge and mass. In both cases the scalar field was chosen
to be massless, i.e. $m^2=0$. The results of \cite{dias2} were extended to the case $m^2=-3$ in 
\cite{brihaye_hartmannNEW}. The qualitative results remain similar.
In the following, we will reinvestigate both $m^2=0$ and $m^2=-3$ and point out further details.

\subsection{$m^2=-3$}

\subsubsection{Charged solitons for $\alpha=0$}

The numerical results demonstrate that for fixed $Q$ the 
solitons exist only for large enough values of $e > e_c$, i.e. exist on an interval
$e\in [e_c(Q):\infty]$. This seems natural when considering
that small $e^2$ corresponds to large $\gamma$ and vice versa. Hence, solitons
exist only if the gravitational coupling is small enough in comparison to the
gauge coupling $e^2$. An estimation of the critical value $e_c(Q)$ for $Q \in [0:10]$
is presented by the solid black line in Fig. \ref{e2_q}. 
\begin{figure}
\centering
\epsfysize=8cm
\mbox{\epsffile{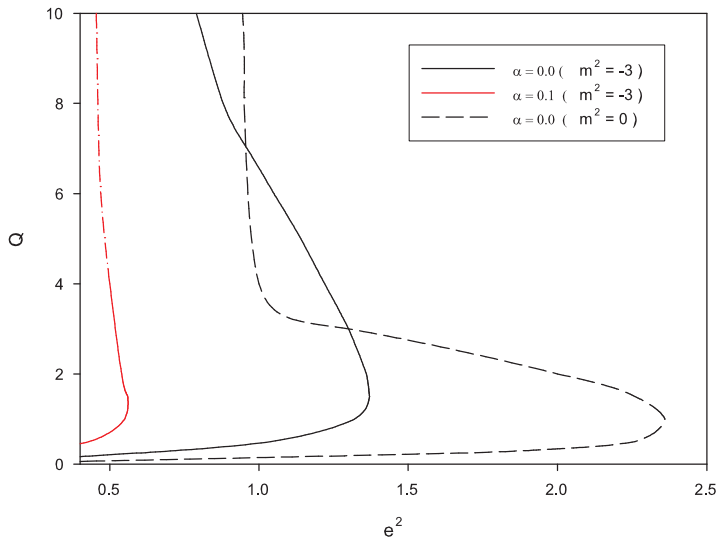}}
\caption{\label{e2_q}
The value of the charge $Q$ as function of the value of the gauge
coupling $e^2$ at which $a(0)=0$ for soliton solutions
with $m^2=0$ and $\alpha=0$ (dashed), $m^2=-3$ and $\alpha=0$ (solid black),
$m^2=-3$ and $\alpha=0.1$ (solid red), respectively. The dotted-dashed red curve is just 
an upper estimate of the critical value $e^2$ since the numerical integration becomes
very difficult here. Note that soliton
solutions exist only to the right, below and above these curves. }
\end{figure}
At the approach of $e_c(Q)$, our numerical results show that
the value of the metric function $a(r)$ at the origin, 
$a(0)$ tends to zero, accordingly the metric becomes singular. 
In Fig.\ref{soliton_a0_e2} the quantity
$a(0)$ is plotted as function of $e^2$ for different values of $Q$. We find in particular
$e_c^2 \approx 1.04$, $1.32$, $1.36$ for $Q=0.5$, $1.0$, $2.0$, respectively.
This can also be seen in Fig. \ref{e2_q}. The mass and the value $\psi(0)$ corresponding to these families 
of solutions are presented in Fig. \ref{soliton_mpsi_e2}. Both $M$ and $\psi(0)$  remain finite for 
$e\rightarrow e_c$. However
we observe that $\psi''(0)$ becomes quite large and likely tends to infinity in this limit.
\begin{figure}[ht]
  \begin{center}
    \subfigure[$a(0)$]{\label{soliton_a0_e2}\includegraphics[scale=0.55]{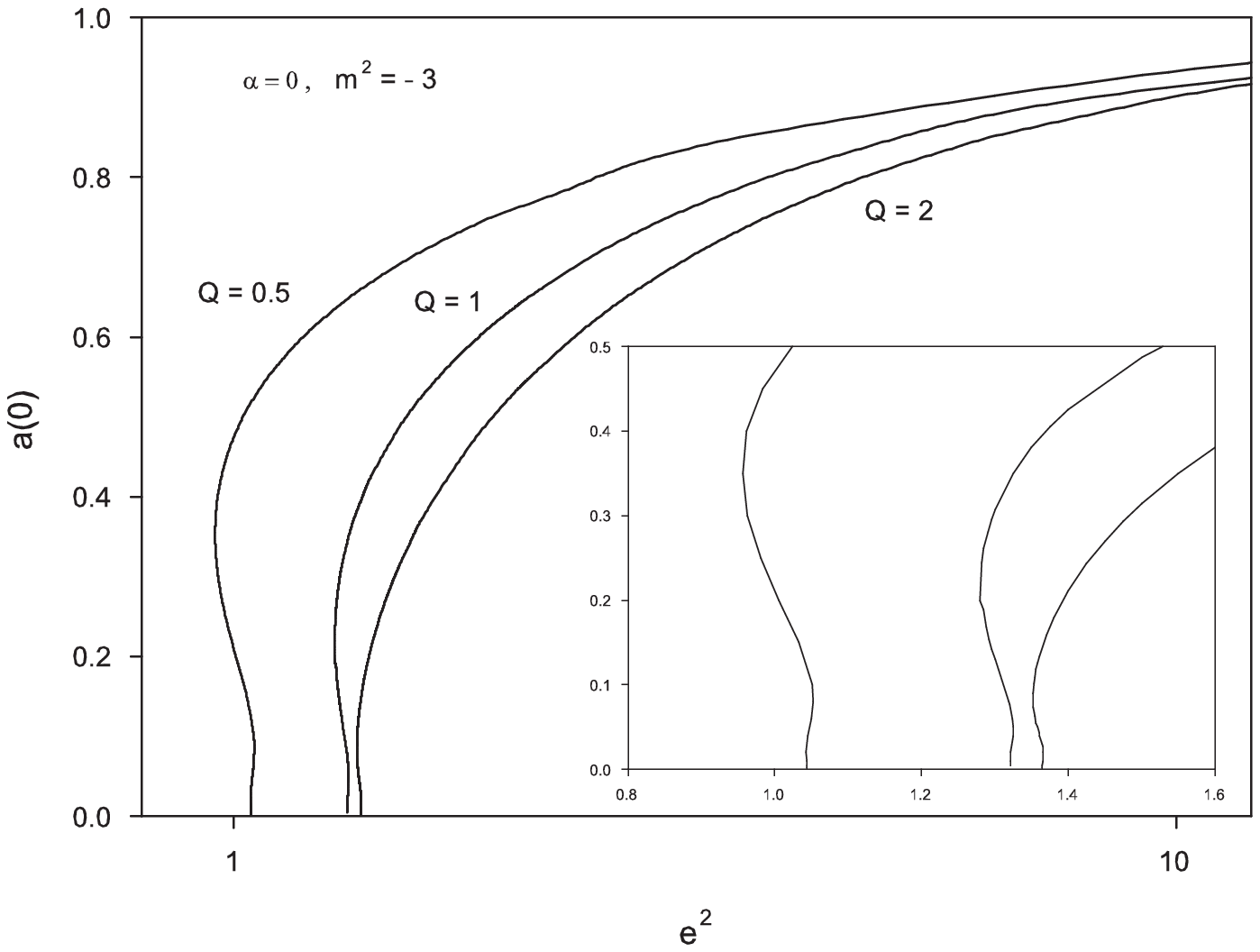}}
    \subfigure[$\psi(0)$ and $M$]{\label{soliton_mpsi_e2}\includegraphics[scale=0.55]{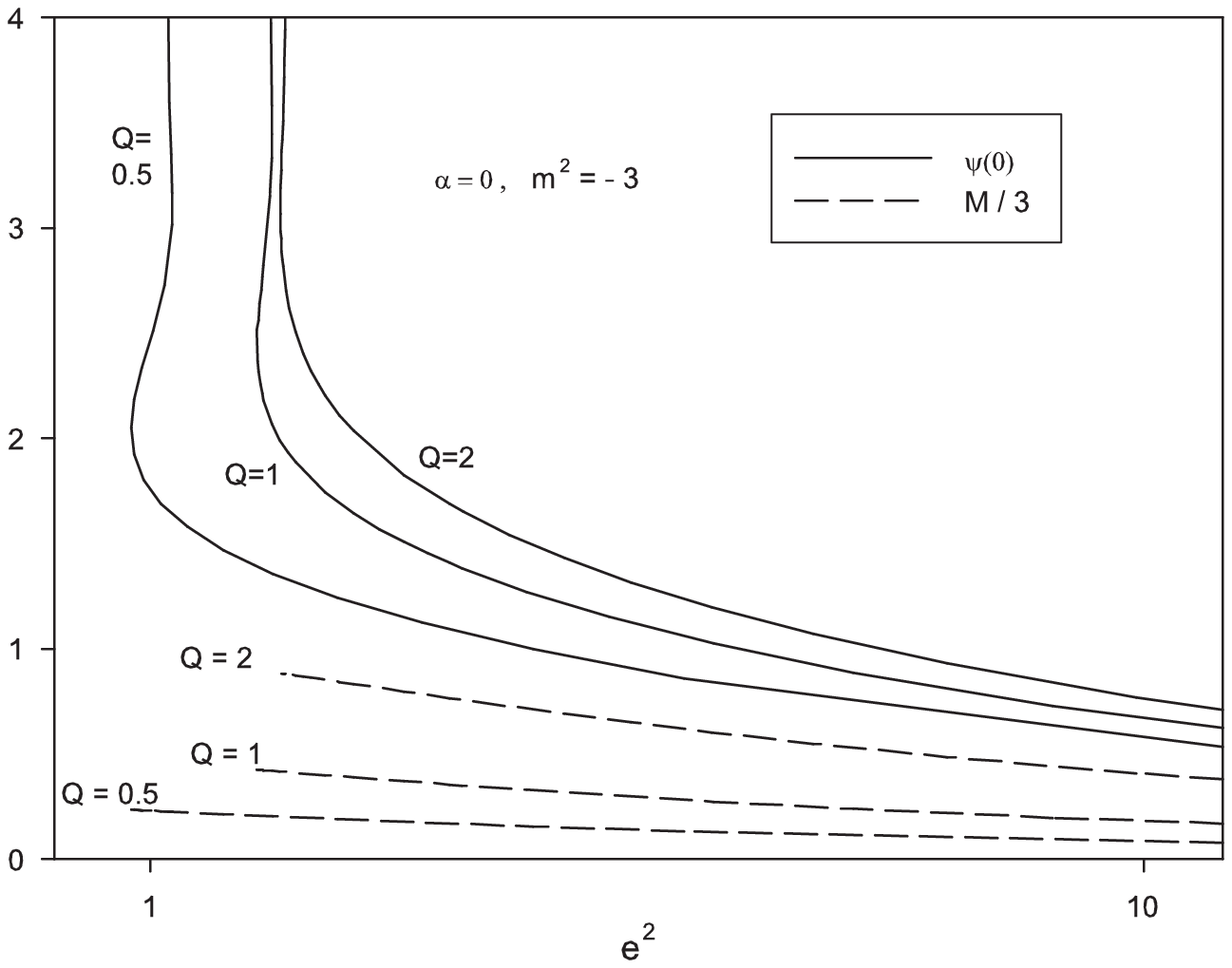}}
    \end{center}
   \caption{The value of the metric function $a(r)$ at the origin, $a(0)$ as function 
of $e^2$ (left) and the value of the scalar field function $\psi(r)$ at the origin, 
$\psi(0)$ and of the
 mass parameter $M$ of the solution as function 
of $e^2$ (right). Here $Q=0.5$, $1.0$, $2.0$, respectively, $\alpha=0$ and $m^2=-3$. }
\label{soliton_e2} 
 \end{figure}


These figures reveal that the  dependence of $a(0)$ on the coupling constant 
$e$ at the approach of the critical value
is not simple because there exist several branches of solutions, i.e.
for a fixed value of $e^2$ more than one soliton  exists and these solutions
are distinguished by their value of $a(0)$. Within our numerical investigations, we 
were able to produce up to three solutions with the same $e$ and different values of $a(0)$
but it might be that more oscillations occur while decreasing $a(0)$ (we investigated
solutions down to $a(0) \sim 10^{-5}$).  Note that this behaviour is also present  in the case
$m^2 = 0$ \cite{dias2}.

Profiles of solitons close to the critical limit  are shown in Fig.\ref{regulier_rh}.
More precisely, the metric and matter functions of the solutions with
 $a(0) = 0.01$ and $a(0)=0.001$, respectively are shown for 
$Q=0.5$ (corresponding to $e_c^2\approx 1.043$).
The functions do not change much with $a(0)$ apart on a small interval close to the origin. 
The metric function $f(r)$ first decreases away from its value at the origin $f(0)=1$.
Then it reaches a minimum while staying nearly constant on a 
plateau, finally it increases to reach its asymptotic behaviour $f(r)\sim r^2$. 
 \begin{figure}
\centering
\epsfysize=8cm
\mbox{\epsffile{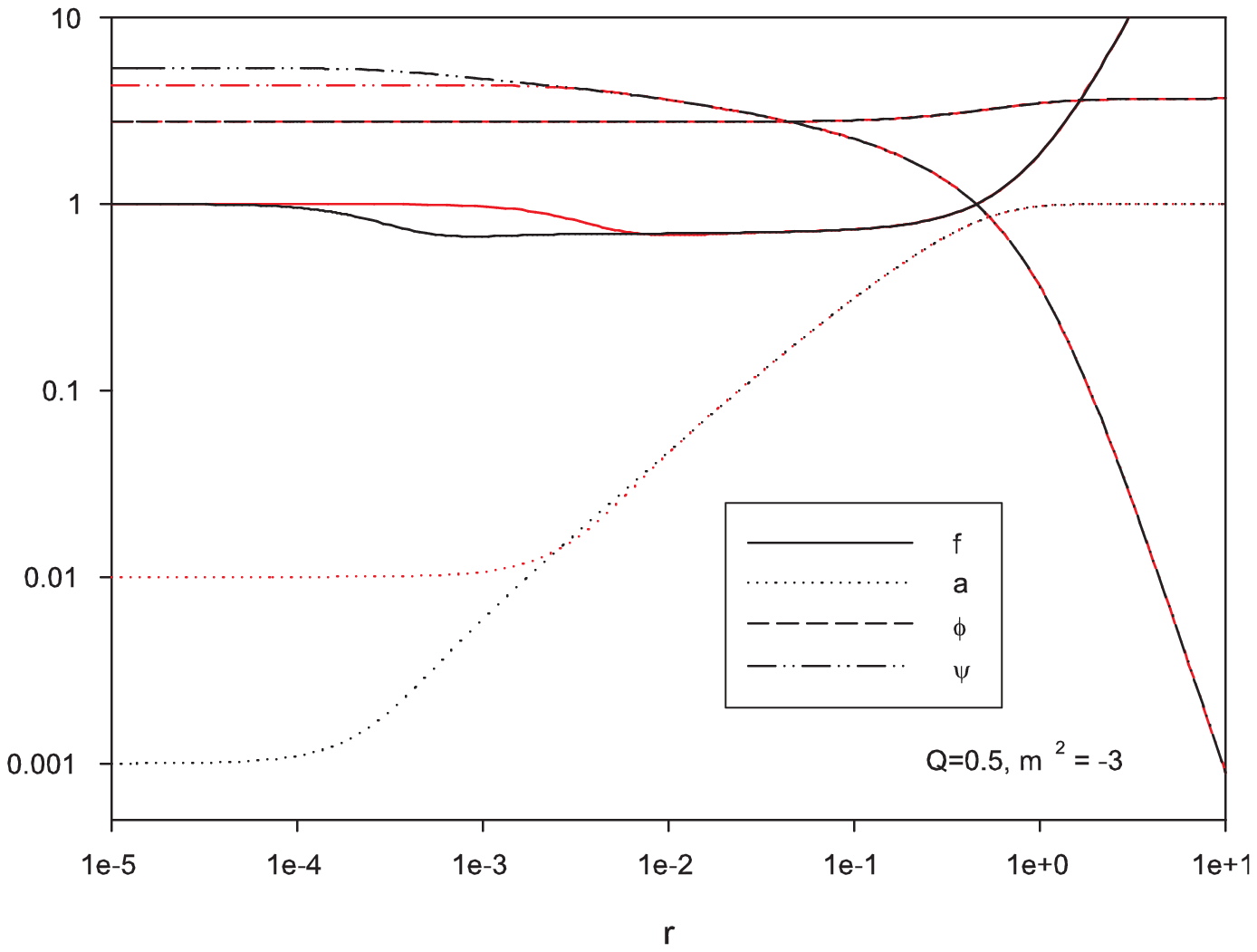}}
\caption{\label{regulier_rh}
The metric and matter functions of a soliton solution with 
$Q=0.5$, $e^2\approx e_c^2$, $\alpha=0$, $m^2=-3$, $a(0) = 
0.01$ (red) and $a(0)=0.001$ (black), respectively.}
\end{figure}

Our numerical results confirm that the solitons can be constructed for arbitrarily 
large values of $e$. In the limit $e \to \infty$ the metric functions  
tend uniformly to $f(r)=1+r^2$ and $a(r)= 1$. 
Note that the limit $e^2 \rightarrow \infty$ corresponds to a decoupling of the
matter from the metric fields, i.e. $\gamma=0$ (the so-called ``probe limit'' in the
context of holographic superconductors).
Hence, we have a fixed global AdS background and two coupled matter field equations 
for $\phi$ and $\psi$ (after suitable rescalings).  

For the purpose of later comparison of the pattern of the solutions for $\alpha=0$ and
$\alpha\neq 0$ (see below)
we present  the dependence of $e^2$, 
of $\phi(0)$ and of $\psi_+$, respectively on $a(0)$ in Fig.\ref{a0_to_0}. Here we
have chosen $Q=0.5$. We observe that all these quantities stay finite at the approach 
$a(0)\rightarrow 0$ if $\alpha=0$ (red curves). We will see in the following that
this is different for $\alpha\neq 0$. 

\begin{figure}
\centering
\epsfysize=8cm
\mbox{\epsffile{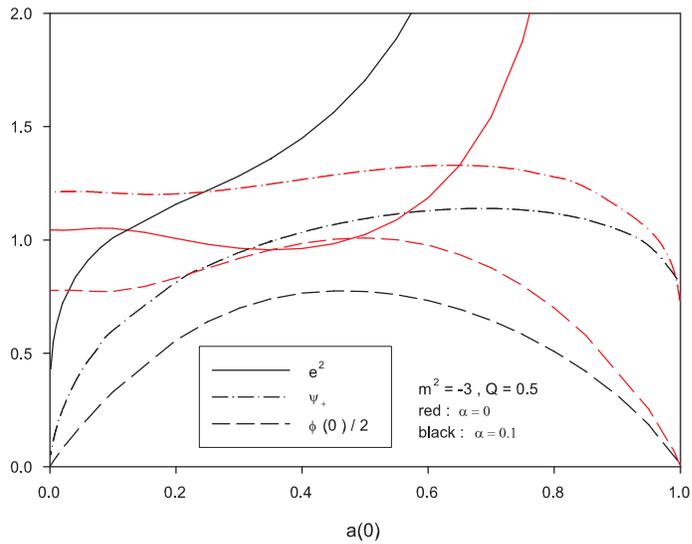}}
\caption{\label{a0_to_0}
We show $e^2$, $\psi_+$ and $\psi(0)$, respectively as function of $a(0)$ for $Q=0.5$ 
for $\alpha=0$
(red) and for $\alpha=0.1$ (black).}
\end{figure}
\subsubsection{Charged Gauss-Bonnet solitons}
The numerical results strongly suggest that the soliton solutions  which exist for $\alpha = 0$
get smoothly deformed when we gradually increase the Gauss-Bonnet coupling constant $\alpha$.
These branches of Gauss-Bonnet solutions exist up to the Chern-Simons limit $\alpha = L^2/4\equiv 1/4$.
Fig.\ref{e2_q} suggests that soliton solutions exist for a larger parameter
range in $Q$ and $e^2$ when choosing $\alpha\neq 0$. Note that solitons
exist to the right, above and below the curves. Hence, the domain in the $e^2-Q-$plane, in which
soliton solutions do not exist is much smaller for $\alpha=0.1$ (red line) 
as compared to $\alpha=0$ (black line). 
In the following, we will describe the properties of these charged Gauss-Bonnet solitons
in more detail.

In Fig.\ref{a0_e2} we give $a(0)$ as a function of $e^2$ for different
values of $\alpha$ and $Q=1$. This is to understand how
the critical behaviour in the limit $a(0)\rightarrow 0$ changes when including Gauss-Bonnet
corrections.
We observe that increasing $\alpha$ from $\alpha = 0$ the value of $e_c$ first increases.
Again, this results in the appearance of several branches of solutions
(up to three for $\alpha=0$ and $\alpha=0.01$). When $\alpha$ is large enough, only one
branch of solutions exists, for our choice of parameters this happens
at $\alpha \gtrsim 0.05$. Increasing $\alpha$ beyond $\alpha=0.05$ the critical
value $e_c$ decreases strongly. Although this pattern was checked only for a few values of $Q$,
we believe that it is generic.

\begin{figure}[ht]
  \begin{center}
    \subfigure[$a(0)$]{\label{a0_e2}\includegraphics[scale=0.55]{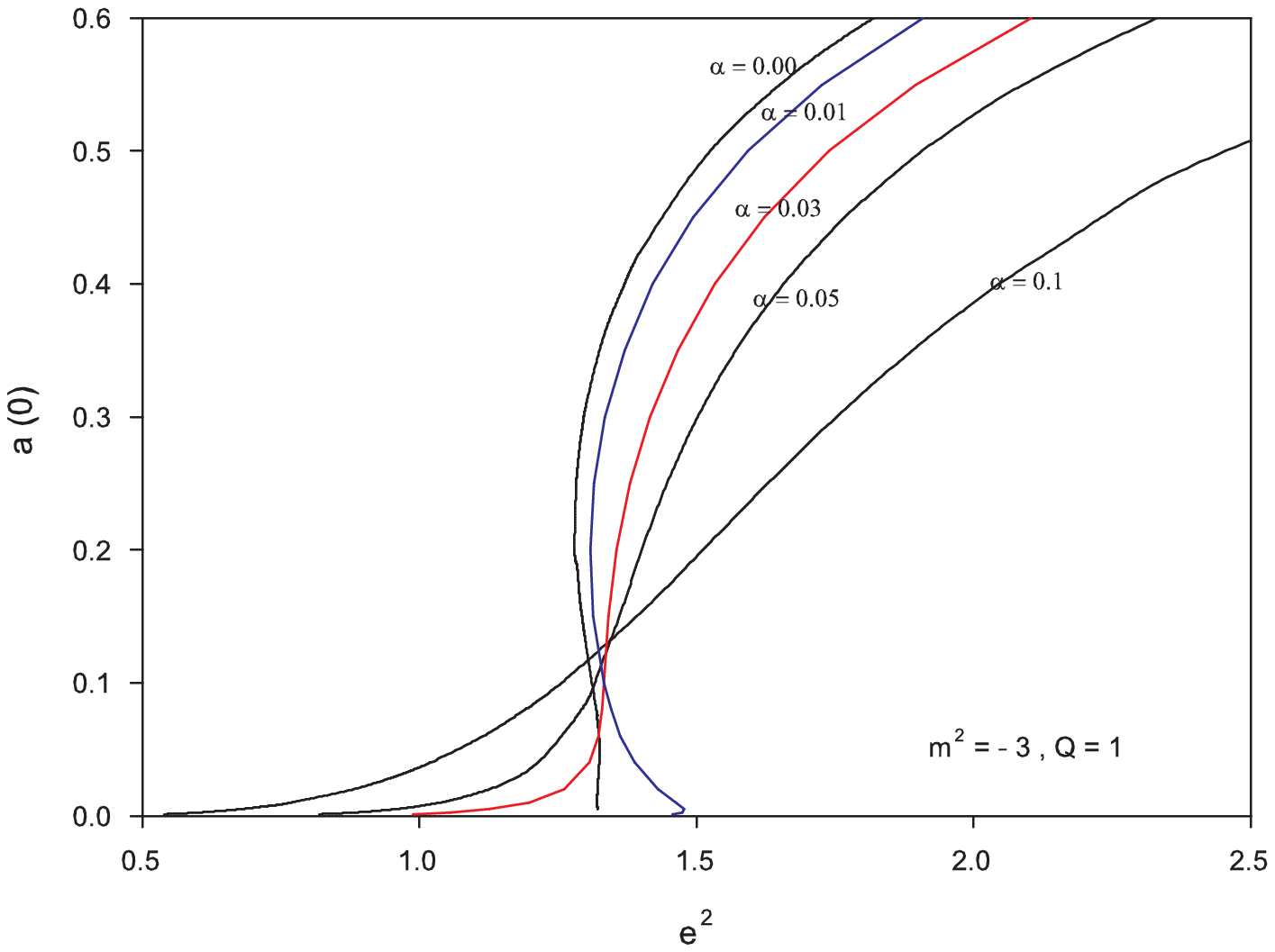}}
    \subfigure[$M$]{\label{mass_e2}\includegraphics[scale=0.55]{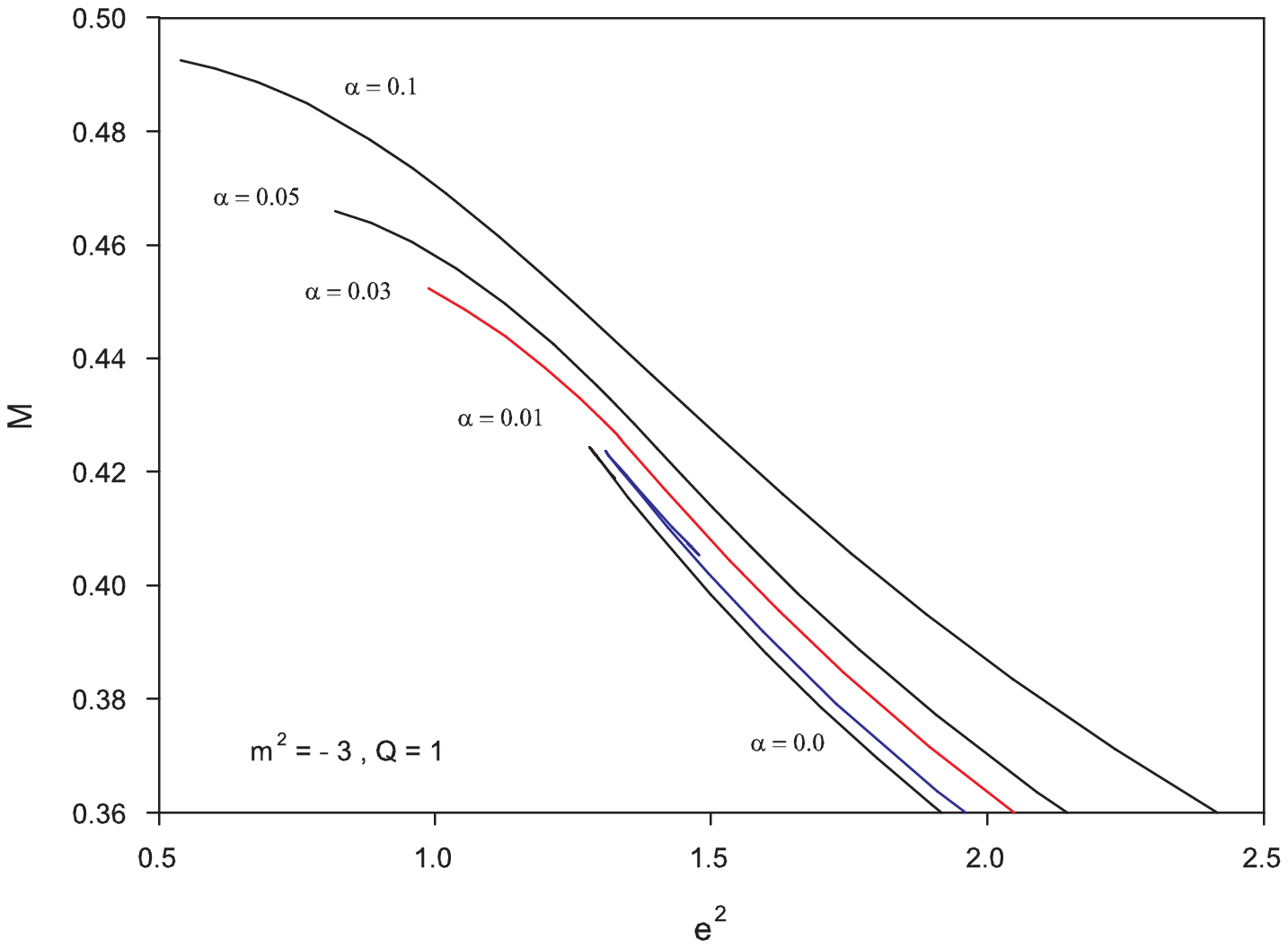}}
    \end{center}
   \caption{The value of the metric function $a(r)$ at the origin, $a(0)$ as function 
of $e^2$ (left) and the value of the
 mass parameter $M$ of the solution as function 
of $e^2$ (right) for several values of $\alpha$. Here $Q=1.0$ and $m^2=-3$.}
\label{e2_alpha} 
 \end{figure}

The masses corresponding to the various branches of Fig. \ref{a0_e2}
are given in Fig. \ref{mass_e2}. 
It turns out that the mass of the soliton increases when increasing the
Gauss-Bonnet parameter $\alpha$ and/or decreasing the gauge coupling $e^2$ such that
the solution with the smallest $e^2$ and largest $\alpha$ has the biggest mass.
Note also that for $\alpha=0$ and $\alpha=0.01$, respectively the three branches of solutions
mentioned above are shown in this figure. The masses of these branches do not differ strongly,
this is why they cannot really be distinguished in the plot. However, we observe that
for a fixed values of $e^2$ the solution on the first branch, i.e. the solution with the 
largest value of $a(0)$ has the lowest mass, while the solution on the third branch, i.e.
the solution with the smallest $a(0)$ has the highest mass. 
This suggests that the solutions on the second and third branch are unstable, 
while the solutions on the  first branch are stable. 

Fig. \ref{a0_e2} suggests further that the approach of the critical
limit $a(0)\rightarrow 0$ is qualitatively different from the case $\alpha=0$. 
This can clearly be seen in Fig.\ref{a0_to_0} where $e^2$, $\phi(0)$ and $\psi_+$ are given
as function of $a(0)$ for $Q=0.5$. As mentioned above,
$e^2$, $\phi(0)$ and $\psi_+$ stay finite for $a(0)\rightarrow 0$ if $\alpha=0$.
For $\alpha=0.1$, on the other hand, the decrease of $a(0)$ results in a 
strong decrease of all three quantities.

To understand this behaviour better, we have looked at the profiles of the
metric and matter functions in this critical limit. 
In Fig. \ref{regulier_rha} we show the 
profiles of Gauss-Bonnet solitons with $\alpha=0.1$ and $Q=0.5$ approaching the 
critical limit $a(0)\rightarrow 0$.
The two solutions correspond to $a(0)=0.01$ ($e^2=0.29$) and $a(0)=0.001$ ($e^2=0.45$), 
respectively.
As can be clearly see in this figure,
the phenomenon observed for $\alpha=0$ (see Fig. \ref{regulier_rh})
exists also here and is in fact enhanced: we find that the metric function
$f(r)$ possesses a very pronounced local minimum which approaches zero for decreasing $a(0)$.
If the value of $f(r)$ at the minimum reached zero before $a(0)$ then the corresponding
limiting solution would correspond to an extremal charged Gauss-Bonnet black hole.
However, our numerical accuracy does not allow us to confirm this statement.
At the moment that the value of $a(0)$ reaches a value that is smaller than our numerical
accuracy $f(r)$ at the minimum has not yet reached zero. We find e.g.
for this choice of parameters that for $a(0) = 0.01$
the minimum is located at $r=r_m\approx 0.317$ with $a(r_m) \approx 0.54$, $f(r_m) = 0.06$
and $\psi(r_m) = 1.7$. For $a(0)=0.001$ we find $r_m\approx 0.309$  and  
$a(r_m) \approx 0.49$, $f(r_m) = 0.01$, $\psi(r_m) = 1.46$. So we cannot make a clear
statement about the nature of the limiting solution yet.
This is a difficult numerical
question that we plan to investigate further in the future.\
However, we are sure that the limiting solution will {\it not} be a extremal
Gauss-Bonnet black hole with scalar hair since it can be proven (see Appendix)
that extremal Gauss-Bonnet black holes do not support scalar hair.
The appearance of the minimum of $f(r)$ is connected to a strong
variation of $e^2$ when decreasing $a(0)$. 
We further observe that $a(r)$ remains roughly constant in the central
region of the soliton and varies strongly in the region where $f(r)$ attains its local minimum.
The pronounced dependence of the electric potential on the value $a(0)$ also contrasts with the
$\alpha=0$ case.
\begin{figure}
\centering
\epsfysize=8cm
\mbox{\epsffile{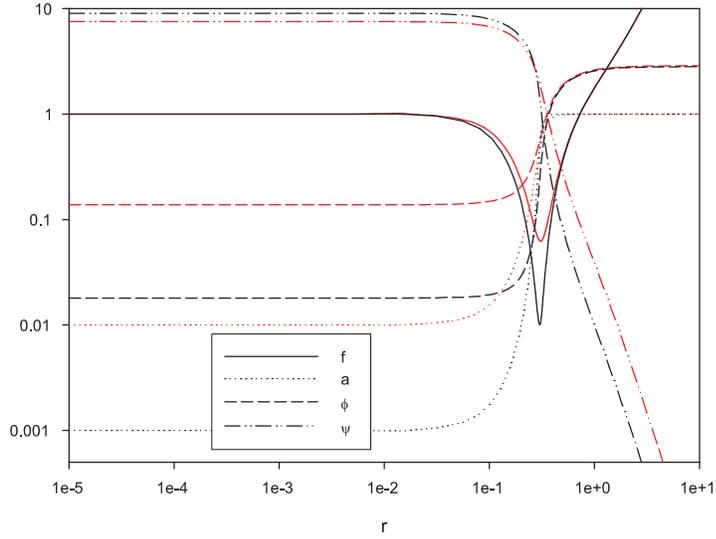}}
\caption{\label{regulier_rha}
The metric and matter functions of a soliton solution with 
$Q=0.5$, $\alpha=0.1$, $m^2=-3$ and $e^2=0.45$ ($a(0) = 0.01$) (red) and 
$e^2=0.29$ ($a(0)=0.001$) (black), respectively. }
\end{figure}

\begin{figure}
\centering
\epsfysize=8cm
\mbox{\epsffile{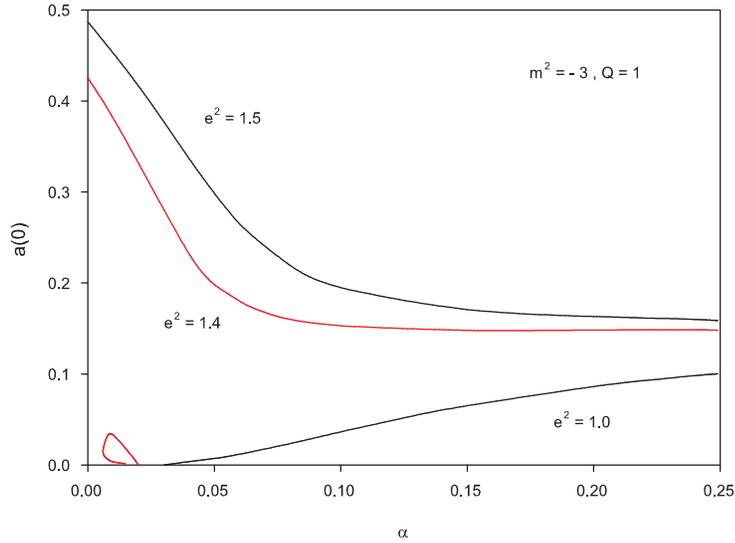}}
\caption{\label{a0_alpha}
The value of the metric function $a(r)$ at the origin, $a(0)$ as function 
of $\alpha$ for $Q=1$, $m^2=-3$ and several values of $e^2$.}
\end{figure}
The observation that for a fixed value of $e^2$ and of $\alpha$ several branches of
solutions can exist  leads to a new type of pattern when plotting $a(0)$ as function
of $\alpha$ for fixed values of $e^2$ and $Q$. This is shown in
Fig. \ref{a0_alpha} for $Q=1$ and three values of $e^2$. 
For small gauge coupling (for our choice of parameters $e^2 \lesssim 1.28$) 
solitons do not exist
for $\alpha = 0$. However, if $e^2$ is large enough 
(here $e^2 \gtrsim 1.28$) solitons exist for the full range of $\alpha \in [0:0.25]$.

For $e^2=1$ the value $a(0)$ is a monotonically increasing function
of $\alpha$ with $a(0)\approx 0.1$ at $\alpha=0.25$ 
and $a(0)=0$ at a critical value $\alpha_c(e^2) \approx 0.03$. 
For $e^2 = 1.5$, on the other hand, the value $a(0)$ is a monotonically decreasing
function of $\alpha$ with $a(0)\approx 0.49$ for $\alpha=0$ and $a(0)\approx 0.16$ for $\alpha=0.25$.
For $e^2 = 1.4$ we observe that there is a 
main branch on which $a(0)$ is monotonically decreasing with $\alpha$
with $a(0)\approx 0.425$ for $\alpha=0$ and $a(0)\approx 0.15$ for $\alpha=0.25$. 
However, there also exists a second branch that is
disconnected from the first and has the form of a small loop that is present
for small values of $\alpha$ and $a(0)$. 
This is in fact a result of the complicated structure of the branches of solutions
that exist for this choice of parameters. This small extra branch
disappears when increasing $e^2$.


\subsection{$m^2=0$}
This case has been studied in detail in \cite{dias2} for $\alpha=0$. 
It was found that for $\tilde{e}^2 \leq \tilde{e}^2_c=\frac{32}{2}$
soliton solutions exist only for small values of $Q$ and that at the maximal
possible value of the charge $Q$ the value of the metric function
$a(r)$ at the origin, $a(0)$ tends to zero. For $\tilde{e}^2 \geq 32/2$ on the other
hand solutions can exist for arbitrarily large $Q$ with $Q$ diverging for $a(0)\rightarrow 0$.
In the following we will demonstrate that for $\tilde{e}^2 \leq \frac{32}{2}$ there exists
in fact a second branch of solutions at large $Q$ which is separated from the first
branch by a ``forbidden band'' of $Q$-values. The existence of a second branch of solutions was already
noticed in \cite{menagerie} for solitons in global 4-dimensional AdS.
Note that our $e^2$ is related to $\tilde{e}^2$ by $e^2=\frac{9}{40} \tilde{e}^2$ such that
$e_c^2= \frac{12}{5}$. 

\subsubsection{Charged solitons for $\alpha=0$}

The domain of existence of solutions in the case $m^2=0$ is given in Fig.\ref{e2_q}
(black dashed line). As already mentioned previously solitons exist 
to the right, above and below this line which itself gives an estimation on 
 the value of the charge $Q$ in dependence on $e^2$ at which
the value of $a(r)$ at the origin vanishes, i.e. where $a(0)=0$. 
For $m^2=0$ and $\alpha=0$ we find that hairy
soliton solutions exist for $Q\in [0:\infty[$ if $e^2 > e^2_c$ - confirming the results stated in \cite{dias2}.
In agreement with \cite{dias2} we find that $e_c^2 \approx 2.4$.

For $e^2_c > e^2 > e^{2,*}_c >0$ on the other hand
soliton solutions exist for $Q\in [0:Q_{f,1}]$ and for $[Q_{f,2}:\infty[$. The difference
between $Q_{f,1}$ and $Q_{f,2}$ becomes larger when decreasing $e^2$. At $e^{2,*}_c$ 
we find that $Q_{f,2}\rightarrow \infty$ such that soliton solutions exist
only for $Q\in [0:Q_{f,1}]$. Our numerical results show that $e^{2,*}_c\approx 0.94$
for $m^2=0$ and $\alpha=0$. 

\begin{figure}[ht]
  \begin{center}
    \subfigure[$Q$ as function of $a(0)$]{\label{a0_q}\includegraphics[scale=0.55]{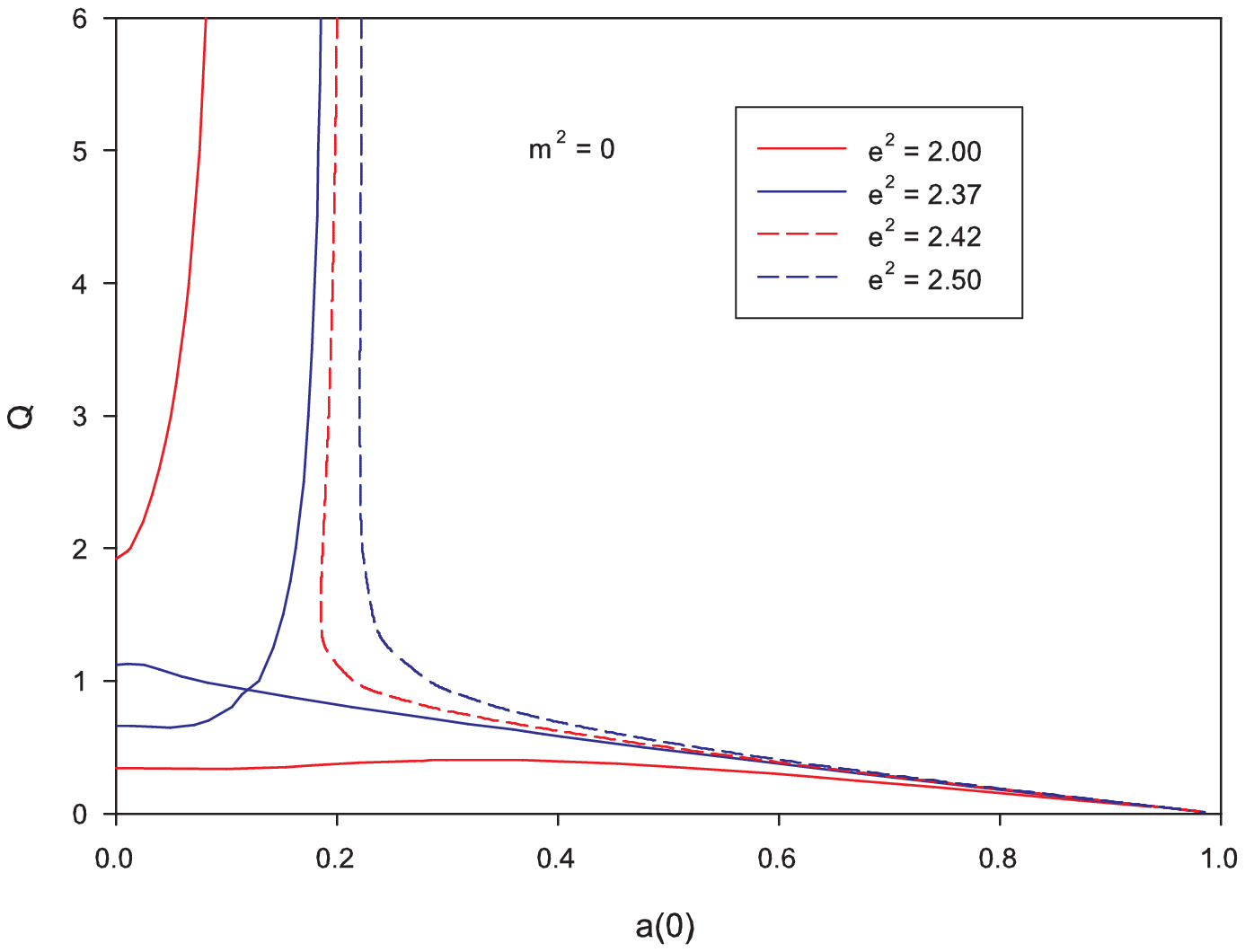}}
    \subfigure[$M$ as function of $\psi(0)$]{\label{psi0_m}\includegraphics[scale=0.55]{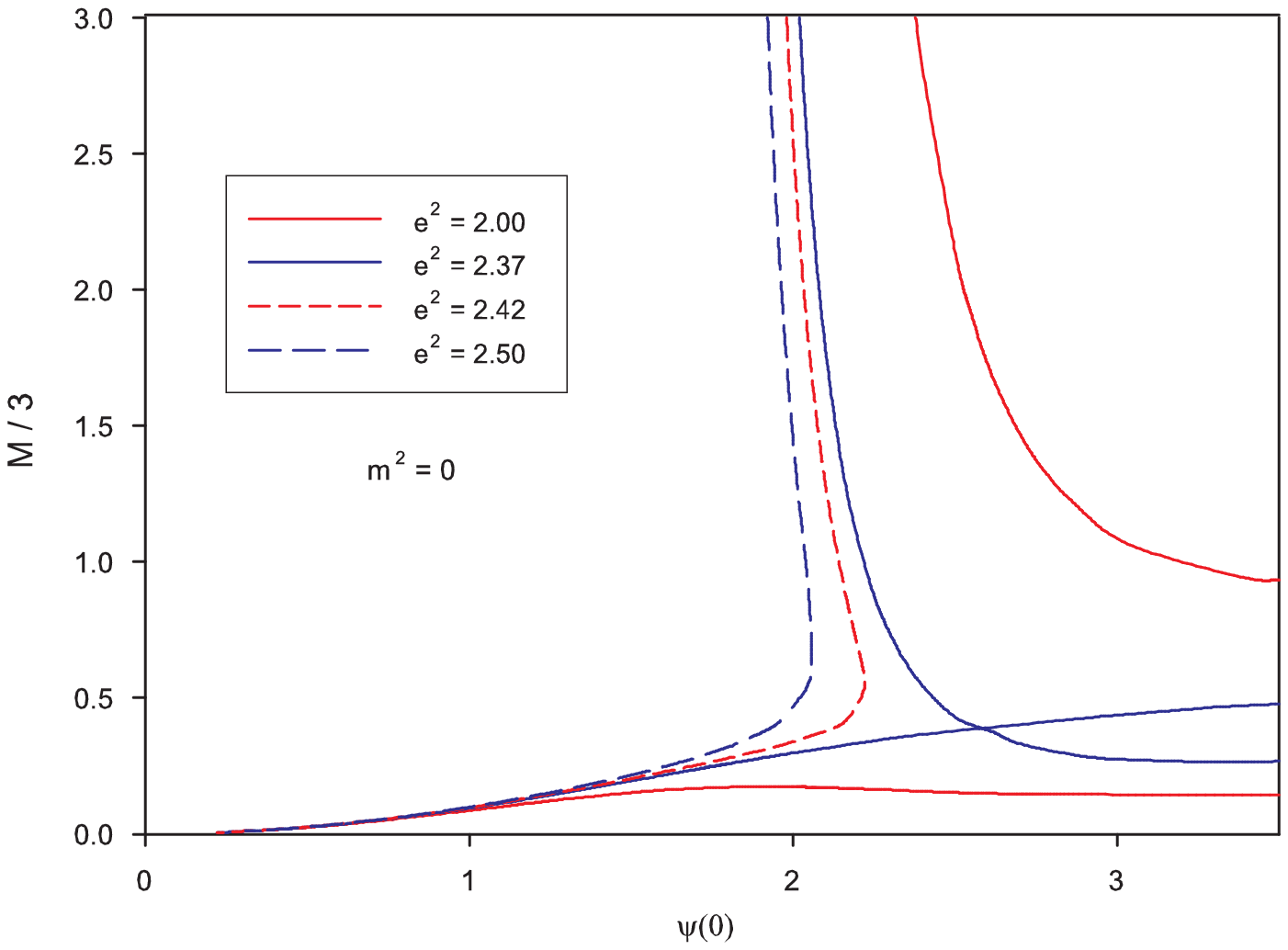}}
    \end{center}
   \caption{The value of the charge $Q$ as function of the value of 
the metric function $a(r)$ at the origin, $a(0)$ (a) as well as the mass $M$ as function of
the value of the scalar field function $\psi(r)$ at the origin, $\psi(0)$ (b)
for soliton solutions
with $m^2=0$, $\alpha=0$, $e^2 =2.0 < e^2_c$, $e^2 =2.37 < e_c^2$, $e^2=2.42 > e_c^2$ 
and $e^2=2.5 > e_c^2$, respectively.}
\label{e2_alpha} 
 \end{figure}

In Fig.\ref{a0_q} we show the value of $Q$ as function of $a(0)$ for $e^2 = 2.0  < e_c^2$,
$e^2=2.37 < e_c^2$, $e^2=2.42 > e_c^2$ and $e^2=2.5 > e_c^2$, respectively. 
As stated above, we find that for $e^2 < e^2_c$
a second branch of solutions exists. For small $e^2$
this is separated from the small $Q$-branch by a ``forbidden band'' 
$Q\in [Q_{f,1}:Q_{f,2}]$
in which soliton solutions do not exist. Increasing $e^2$ these two branches approach each other 
such that at some stage they intersect for some given value of $a(0)$. This is seen clearly for $e^2=2.37$.
Increasing $e^2$ even further such that $e^2 \rightarrow e_c^2$ we find that the two
branches start to merge and become one continuous curve for $e^2 > e^2_c$. For $e^2 > e^2_c$ there are no solutions
for $a(0)=0$.   
In Fig.\ref{psi0_m} we show the corresponding mass $M$ of the solutions in dependence
on $\psi(0)$. These curves are qualitatively similar to the curves in 4-dimensional AdS given in \cite{menagerie}. 
The $M=Q=0$ limit corresponds to global AdS with radius $L_{\rm eff}$. This corresponds to $a(0)=1$ in the $Q$-plot
and $\psi(0)=0$ in the $M$-plot. Moreover, we observe that solutions
with $Q$ and $M$ tending to infinity are possible. These are the planar solutions mentioned previously.
Moreover, for $\psi(0)$ becoming large (and equivalently $a(0)$ tending to zero for $e^2 < e^2_c$ and
$a(0)$ tending to its critical, non-vanishing value for $e^2 > e^2_c$), 
the mass of the solutions tends to a constant.

The critical behaviour at the approach $a(0) \to 0$ was discussed
in \cite{dias2} and is qualitatively similar to the case $m^2=-3$ discussed in this paper. 
In Fig.\ref{soliton} we show the dependence of $a(0)$ on $e^2$ for
different values of $Q$ (black lines).

\subsubsection{Charged Gauss-Bonnet solitons}

We observe that the dependence of $a(0)$ on $e^2$ looks qualitatively similar 
to the case $m^2=-3$. 
This can be seen in Fig.\ref{soliton}, where we show
$a(0)$ as function of $e^2$ for  several values of $\alpha$ and of $Q$. 
Increasing $\alpha$ from $\alpha = 0$ we observe that the oscillations
around the critical value $e_c$ are first amplified 
(see red curve in Fig.\ref{soliton} for $\alpha=0.01$  and $Q=0.5$). 
This results in  the existence  of several solutions with different values of 
$a(0)$ for the same value of $e^2$. E.g. we were 
able to construct up to three solutions for $\alpha=0$ and $\alpha=0.01$, respectively.
Increasing $\alpha$ further, only one
branch of solutions exists. For our choice of parameters this happens
at $\alpha \gtrsim 0.05$. Increasing $\alpha$ beyond $\alpha=0.05$ the critical
value $e_c$ decreases strongly while increasing $\alpha$.  

\begin{figure}
\centering
\epsfysize=8cm
\mbox{\epsffile{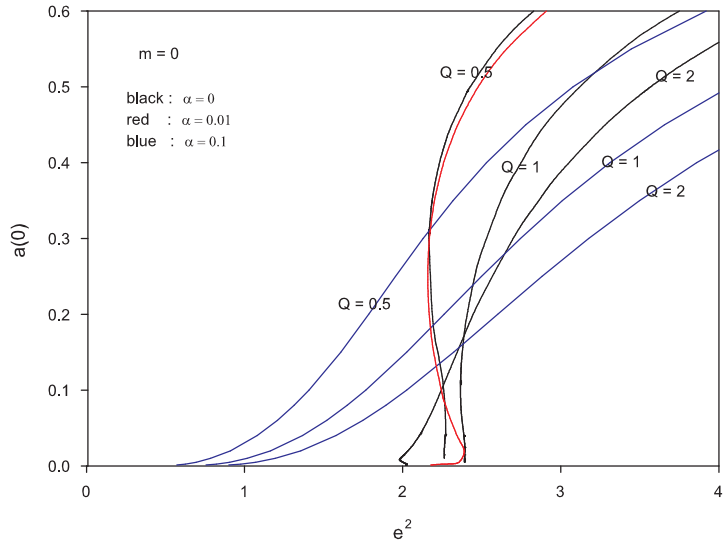}}
\caption{\label{soliton}
The value of the metric function $a(r)$ at the origin, $a(0)$ as function of $e^2$ for a  
massless scalar field
and different values of $Q$ and $\alpha$.
}
\end{figure}

%
%
\section{Black holes}

In order to find an explicit black hole solution of the equations of motion, we have
to fix appropriate boundary conditions. 
In the following, we are interested in the formation of scalar hair on
electrically charged black holes with a regular horizon at $r=r_h$ such that
\begin{equation}
\label{bc1}
 f(r_h)=0  
\end{equation}
with $a(r_h)$ finite. 
In order for the matter fields to be regular at the horizon we need to impose:
\begin{equation}
\label{bc2}
\phi(r_h)=0 \ \ , \ \ \psi'(r_h) = \left.\frac{ m^2 \psi\left(r^2+2\alpha k\right)}  {2r + 4r/L^2 -\gamma r^3 \left( m^2 \psi^2 +  \phi'^2/(2a^2)\right)   }\right\vert_{r=r_h}   \ .
\end{equation}
Black hole solutions can therefore be 
characterized by the parameters $r_h$, $Q$, $e^2$ as well as the Gauss-Bonnet coupling $\alpha$.

The Hawking temperature of these
black holes reads
\begin{equation}
          T_{\rm H} = \left.\frac{1}{4\pi}\sqrt{-g^{tt} g^{KL} \partial_K g_{tt} 
\partial_L g_{tt}}\right\vert_{r=r_h}= \frac{1}{4 \pi} f'(r_h) a(r_h)  \ , \ \ K,L=1,2,3,4  \ ,
\end{equation}
while the entropy is
\begin{equation}
 S=\frac{\pi^2}{2G}\left(r_h^3 + 6\alpha r_h\right)  \ .
\end{equation}

The stability of black hole solutions for $\alpha=0$ and $m^2=0$
was studied in \cite{dias2}. It was found that RNAdS black holes
are never unstable to scalar condensation for $\tilde{e}^2 \leq 3$ and unstable
for all charges $Q$ for $\tilde{e}^2 \geq \frac{32}{3}$. In the intermediate
interval for $3 \leq \tilde{e}^2 \leq  \frac{32}{3}$ RNAdS black holes become
unstable for  sufficiently large charge $Q$, i.e. for $Q \geq Q_c(\tilde{e}^2)$.
Moreover, it was observed that for large gauge coupling and
small charges the solutions exist all the way down to vanishing horizon. The limiting
solutions are the soliton solutions mentioned above. On the other hand for large charge the
limiting solution is a singular solution with vanishing temperature and finite entropy, which
is not a regular extremal black hole \cite{fiol1}. Similar results were obtained for
a tachyonic scalar field with $m^2=-3$ in \cite{brihaye_hartmannNEW}. Here we reinvestigate
both cases and point out further details.

\subsection{$m^2=-3$}

\subsubsection{Hairy charged black holes for $\alpha=0$}

\begin{figure}
\centering
\epsfysize=7cm
\mbox{\epsffile{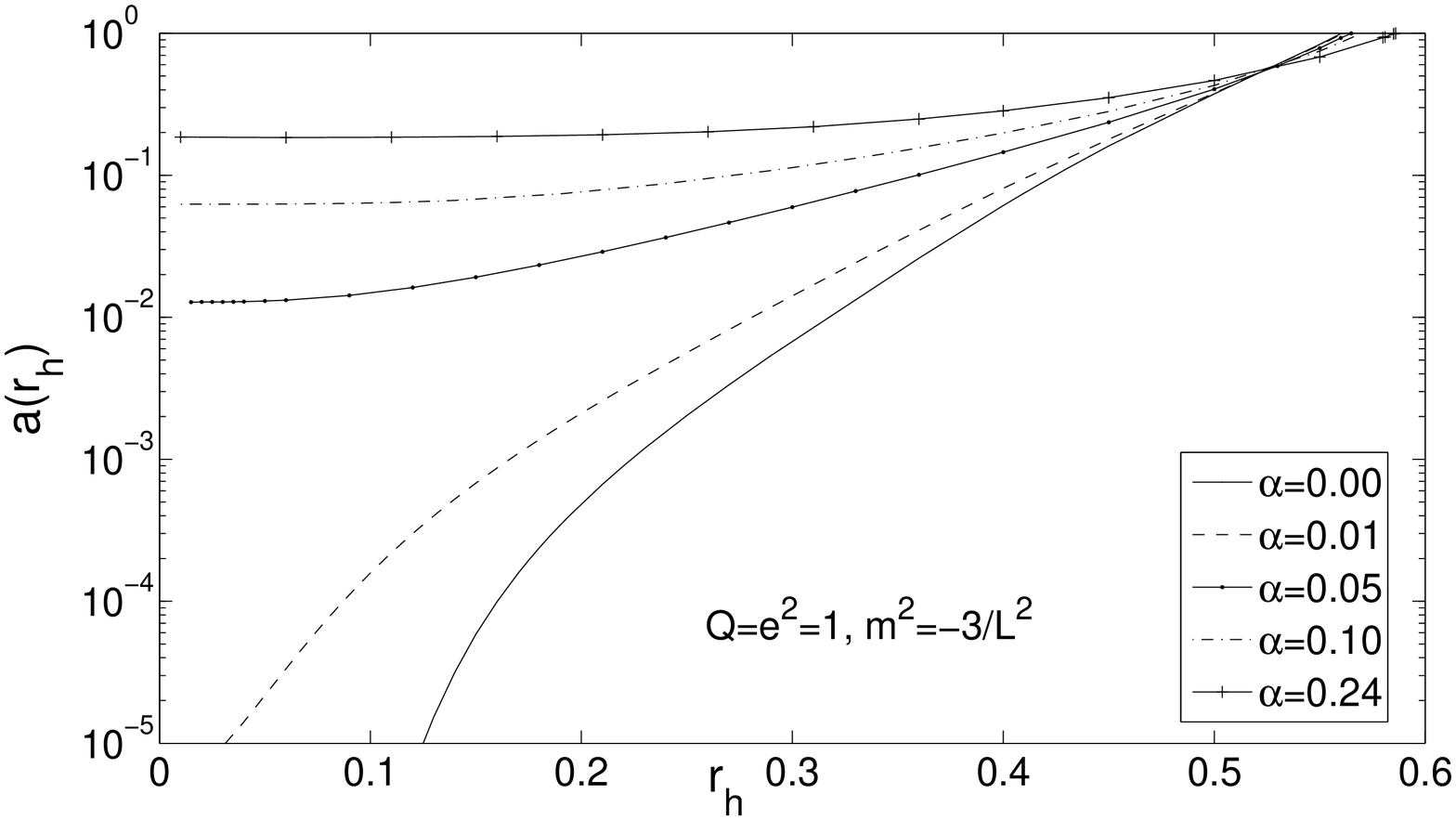}}
\caption{\label{a_xh}
The value of $a(r_h)$ as function of $r_h$ for $m^2=-3/L^2\equiv -3$, $e^2=1$, $Q=1$ and different values of the Gauss-Bonnet coupling $\alpha$.}
\end{figure}

We have first studied the interval of $r_h$ on which black holes exist
for $Q=1=e^2$. This is shown in 
Fig.\ref{a_xh}, where we give $a(r_h)$ as function
of $r_h$.
At the maximal value of $r_h$, $r_{h,max}$ the solution tends to a
black hole solution without scalar hair, i.e. the RNAdS solutions with 
$a(r)\equiv 1$ and $\psi(r)\equiv 0$. The value $r_{h,max}$ depends
on $e$ and $Q$, e.g. for $Q=1$ we have  $r_{h,max} = 0.56, 0.59, 0.62$, respectively for 
$e^2=1.0,1.5,2.0$.

At the minimal value of $r_h$, $r_{h,min}$ the solutions either tend to
a singular black hole solution with $a(r_{h,min})=0$ and correspondingly temperature 
$T_{\rm H}=0$ or to the soliton solutions mentioned early in this paper. For fixed
$Q$ this depends on the choice of $e^2$. This
can been see in Fig.\ref{a_xh} and Fig.\ref{comparaison_1}. The curve of $a(0)$
tends to zero for $r_h >0$ when choosing $Q=1=e^2$, $\alpha=0$ (see Fig.\ref{a_xh}).
Increasing $e^2$ this changes.
For $e^2$ large enough the
black holes exist all the way down to $r_h=0$, where they merge with the
soliton solutions discussed above. 
Generally, we can say that for a fixed $Q$ the black holes tend to the
soliton solutions for $e^2 \geq e_c^2$ (i.e. for values of $e^2$ for which
soliton solutions exist), while they tend to the singular black hole solutions
for $e^2 < e_c^2$. For $Q=1$ we have found (see discussion on solitons above) that
$e_c^2\approx 1.32$. Hence we would expect that the black holes merge with the
solitons for $e^2 \geq 1.32$. This is indeed what our numerics shows and is
given for $Q=1$ and $e^2=2$
in Fig. \ref{comparaison_1}. Here, we plot the matter and metric functions
corresponding to a black hole with small $r_h=0.01$. 
Clearly, the black hole solution approaches the
soliton for $r_h\rightarrow 0$. While $a(r)$ and $\psi(r)$ are nearly identical on the
full $r$-interval, the behaviour of $f(r)$ and $\phi(r)$ close to the origin
differs between the black hole and the soliton case. This is not surprising, since
the boundary conditions for black holes are $f(r_h)=0$, $\phi(r_h)=0$, while
for solitons we choose $f(0)=1$ and $\phi(0)=\phi_0 > 0$.
Fig.\ref{comparaison_1} also suggests that $f'(r_h)$ and $\phi'(r_h)$ tend to infinity
for $r_h\rightarrow 0$. 
This is shown in Fig.\ref{gauss_bonnet_phys_1}, where we give
the values of $\psi(r_h)$, $f'(r_h)$ and $\phi'(r_h)$, respectively
as function of $r_h$ for $e^2=2$ and $Q=1$ (black lines).
Clearly, both $f'(r_h)$ and $\phi'(r_h)$ tend to infinity.
The value $\psi(r_h)$ reaches a finite value for $r_h\rightarrow 0$, e.g.
$\psi(r_h\rightarrow 0)\rightarrow 1.4$ for $e^2=2$, $Q=1$.

\begin{figure}
\centering
\epsfysize=8cm
\mbox{\epsffile{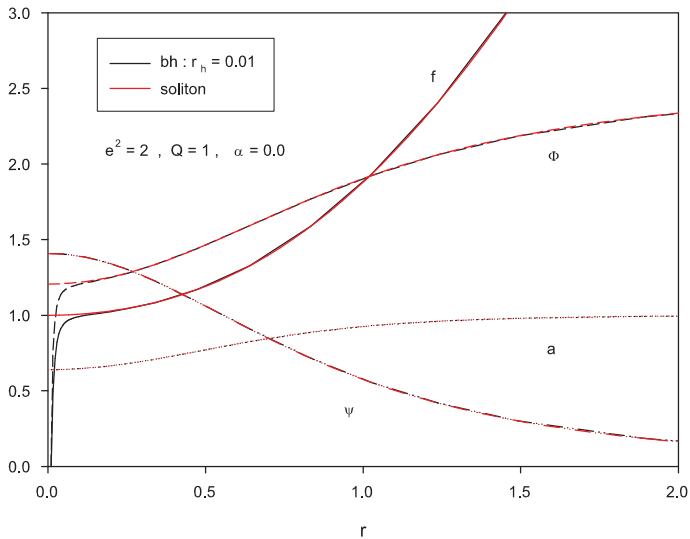}}
\caption{\label{comparaison_1}
The matter and metric functions of a black hole solution
with $r_h=0.01$ and the corresponding soliton solution for $Q=1$, $e^2=2$.}
\end{figure}

\begin{figure}
\centering
\epsfysize=8cm
\mbox{\epsffile{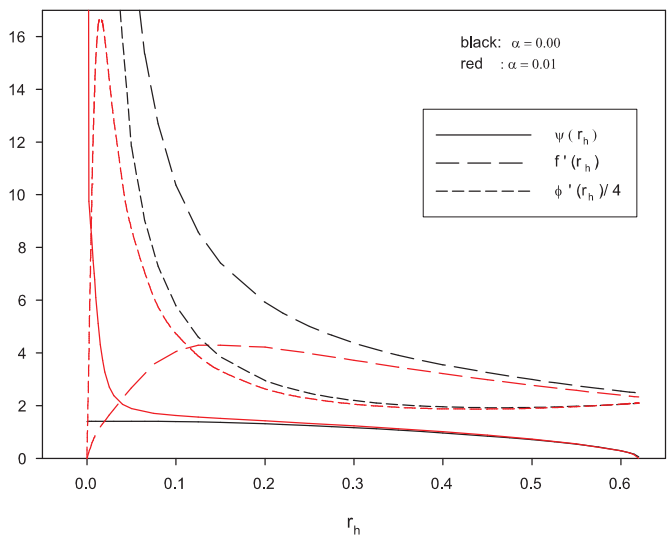}}
\caption{\label{gauss_bonnet_phys_1}
The values of $\psi(r_h)$, $f'(r_h)$ and $\phi'(r_h)$, respectively as function 
of $r_h$ for $\alpha=0$ (black) and $\alpha = 0.01$ (red). Here $e^2=2$, $Q=1$.}
\end{figure}

We have also studied the behaviour of the solutions for fixed $r_h$ 
and varying $e^2$. We find that the black holes exist
for large enough values of
$e^2$. In the limit $e^2\to \infty$ the scalar field function slowly 
approaches the null function. However,
we observe a new phenomenon: we find that the metric
function $f(r)$ is no longer monotonically increasing and develops a local 
minimum at $r_m$ where $r_h > r_m > \infty$ for sufficiently small $e^2$. 
This is shown in Fig. \ref{new_lue_1} for $r_h=0.525$, $Q=1$ and $e^2=0.6975$.
Decreasing $e^2$ the value of the local minimum of $f(r)$ decreases further and
the solution evolves into a configuration
with different  features for $r \in [r_h,r_m]$ and $r \in [r_m,\infty]$, respectively.
For $r \in [r_h,r_m]$ the scalar field function $\psi(r)$ is clearly non-vanishing
and the value $a(r_h)$ becomes
very small - in fact of order $10^{-6}$. This renders the numerical construction 
quite involved and it becomes unreliable at some stage because 
the value $a(r_h)$ gets smaller than the
accuracy of our numerical code. On $[r_m,\infty]$ we have $a(r) \sim 1$ 
and $\psi(r) << 1$. 

\begin{figure}
\centering
\epsfysize=8cm
\mbox{\epsffile{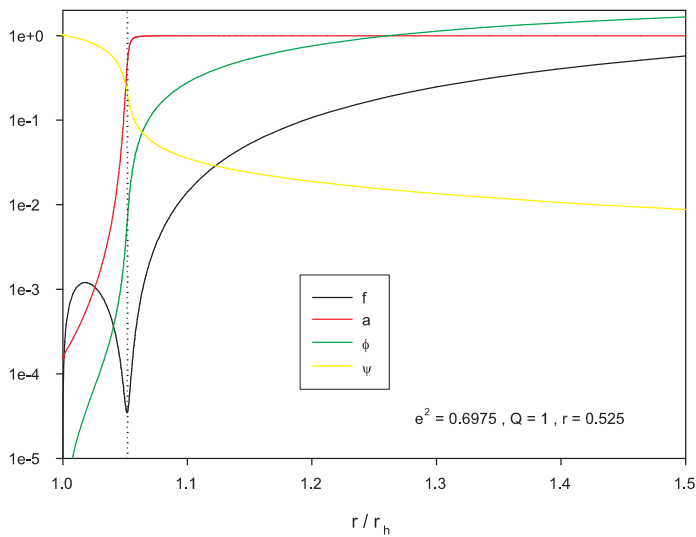}}
\caption{\label{new_lue_1}
The metric and matter functions of a hairy RNAdS solution with $r_h=0.525$, $Q=1$ 
and  $e^2 = 0.6975$ are given.}
\end{figure}

The occurrence of this phenomenon depends crucially on the
fact that a RNAdS black hole exists for the chosen values of $Q$ and $r_h$.
RNAdS black holes with horizon at $r=r_h$ indeed exist for small $Q$ and become extremal for
$Q=Q_m$ with
$Q_m^2 = r_h^4(1+r_h^2)/\gamma$. When considering hairy black holes with $Q$ and $r_h$
we find the following when decreasing $e$: (a) if a RNAdS solution with the same $Q$ and $r_h$ exists
then the hairy black holes tend to the corresponding
   RNAdS black hole for a finite value of $e$, 
(b) if a RNAdS black with the same $Q$ and $r_h$ does {\it not}
exist an additional local minimum develops and the above described phenomenon
appears. 
\subsubsection{Hairy charged Gauss-Bonnet black holes}
Our numerical results suggest that the black holes which exist for $\alpha=0$
get continuously
deformed for $ 0 < \alpha < L^2/4\equiv 1/4$. In the following, we want to discuss in which way
these charged Gauss-Bonnet black holes differ from the charged black hole
solutions in the $\alpha=0$ limit.

We have first studied the interval of the horizon radius $r_h$, $[r_{h,min}: r_{h,max}]$ 
on which the black hole solutions
exist for fixed $Q$, $e^2$ and $\alpha$. 
At the maximal value of $r_h$, $r_{h,max}$ the solution tends to a
black hole solution without scalar hair, in which case $\psi(r)\equiv 0$ and the remaining 
functions are
given by (\ref{rn}). 

At the minimal value of $r_h$, $r_{h,min}$ the solutions either tend to
a singular black hole solution with $a(r_{h,min})=0$ or to a singular solution
with $f´(r_h)\rightarrow 0$ for $r_{h}\rightarrow 0$, while $a(r_h)$ stays finite.
We will demonstrate in the following that the limiting solution for $r_h\rightarrow 0$ is 
not the Gauss-Bonnet soliton solution discussed above. 

In Fig.\ref{a_xh} we give $a(r_h)$ as function
of $r_h$ for different values of $\alpha$ for $Q=1$ and $e^2=1$.
We observe that $r_{h,max}$ increases slightly with $\alpha$. 
Moreover, for $Q=e^2=1$ there is no
soliton solution for $\alpha=0$ and the black hole solutions
end at a singular black hole solution. This however changes when
increasing $\alpha$. For $\alpha$ sufficiently large (for our choice of 
parameters for $\alpha \gtrsim 0.01$) there exist
black hole solutions for $r_h \in ] 0: r_{h,max}]$.
This is clearly seen in Fig.\ref{a_xh}. 

One important observation is that the limit $r_h\rightarrow 0$ is 
different as compared to the $\alpha=0$ case. 
This is shown in
Fig.\ref{comparaison_2} where we give the matter and metric functions of a Gauss-Bonnet
black hole with $r_h=0.01$ and of the corresponding soliton solution for $e^2=Q=1$ and
$\alpha=0.24$. 
Rather then tending to the soliton, the black hole approaches a singular configuration 
for $r_h \to 0$. To strengthen this claim we show 
$f'(r_h)$, $\phi'(r_h)$ and $\psi(r_h)$, respectively, as function of $r_h$ in Fig. \ref{gauss_bonnet_phys_1}.
For $\alpha\neq 0$ we find that
the values $f'(r_h)$, $\phi'(r_h)$ smoothly reach zero for $r_h \to 0$ while
$\psi(r_h)$ diverges. 
As a consequence, the Gauss-Bonnet black holes do not approach the Gauss-Bonnet solitons
discussed above in the limit $r_h \to 0$, which have $f(0)=1$ and $\phi(0)=\phi_0 > 0$. 
This is completely different then in the $\alpha=0$ case where $f'(r_h)$ and $\phi'(r_h)$ both tend
to infinity for $r_h\rightarrow 0$.

The difference is also seen when comparing the entropy, temperature and mass of the limiting
solutions. For $\alpha=0$ there are two different limits: if the black hole approaches the
soliton, the limiting solution has infinite temperature, while the entropy tends to zero.
On the other hand, if the black hole tends to a singular black hole solution the temperature tends
to zero, while the entropy stays finite.
For $\alpha > 0$ the temperature also seems to tend to zero in the limiting case.
Independent of the choice of $\alpha$ the mass remains  finite.

\begin{figure}
\centering
\epsfysize=8cm
\mbox{\epsffile{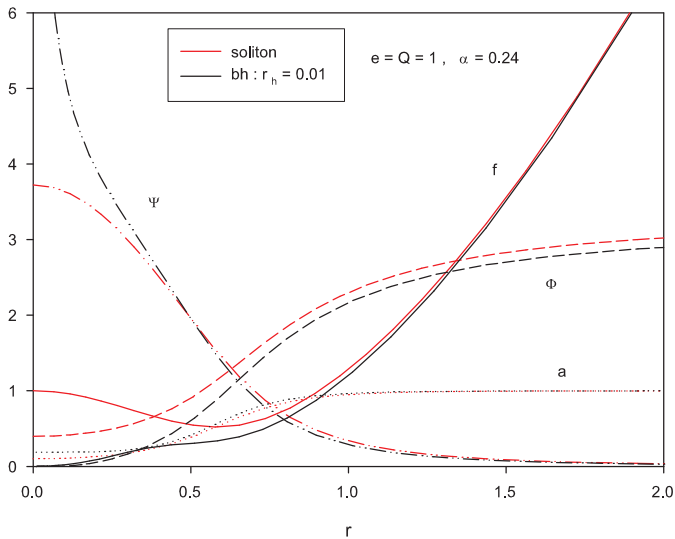}}
\caption{\label{comparaison_2}
The matter and metric functions of the Gauss-Bonnet black hole with $r_h=0.01$ as well
as those of the corresponding soliton 
with $e^2=Q=1$ and $\alpha =0.24$.}
\end{figure}
We also observe that $\psi_+$ remains finite 
and positive in the limit $r_h  \to 0$.
The effect of $\alpha$ on the quantities given in
Fig. \ref{gauss_bonnet_phys_1}  is shown in Fig.\ref{gauss_bonnet_phys_1_b}.
Complementary to the figures above, the  dependence on $\alpha$ of the same parameters
characterizing the solutions is shown in Fig. \ref{gauss_bonnet_phys_2}.

\begin{figure}[ht]
  \begin{center}
    \subfigure[$\alpha$ fixed]{\label{gauss_bonnet_phys_1_b}\includegraphics[scale=0.55]{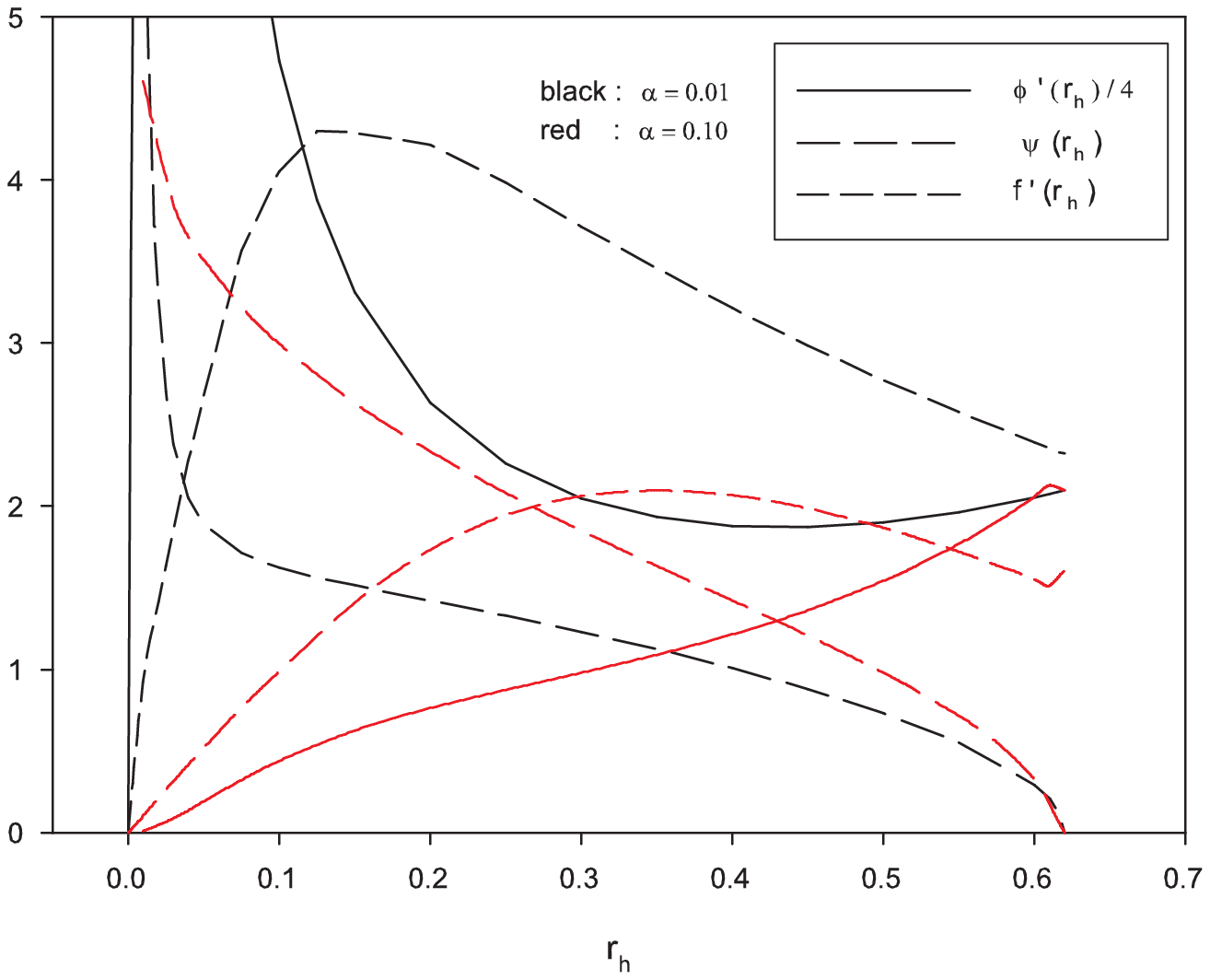}}
    \subfigure[$r_h$ fixed]{\label{gauss_bonnet_phys_2}\includegraphics[scale=0.55]{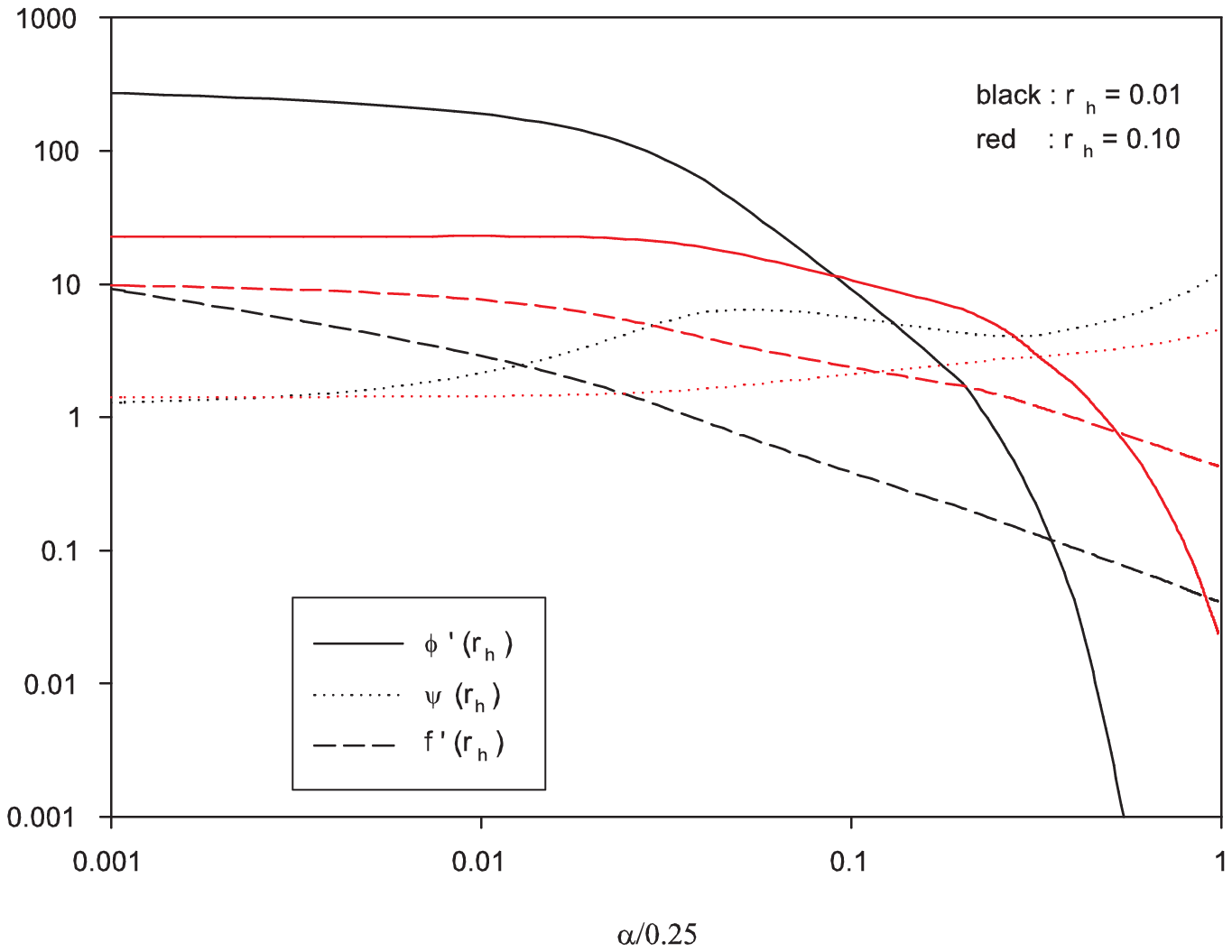}}
    \end{center}
   \caption{The values of $\psi(r_h)$, $f'(r_h)$ and $\phi'(r_h)$ as function 
of $r_h$ for two different values of $\alpha$ (left) and
as function 
of $\alpha$ for different values of $r_h$ (right). Here $e^2=2$, $Q=1$. }
\label{gauss_bonnet_phys} 
 \end{figure}


The behaviour of the black hole occurring while the coupling constant $e^2$ decreases
seems to persist for $\alpha >0$.
This is illustrated in Fig. \ref{gauss_bonnet_bh_3}.
The numerical results suggest that a local minimum of the metric function $f(r)$ forms
that seems to tend to zero. Our numerical investigation doesn't allow
as to make a clear statement about the limiting solution. However, given the
prove in the Appendix, we are sure that this is {\it not} an extremal Gauss-Bonnet
black hole with scalar hair.

\begin{figure}
\centering
\epsfysize=8cm
\mbox{\epsffile{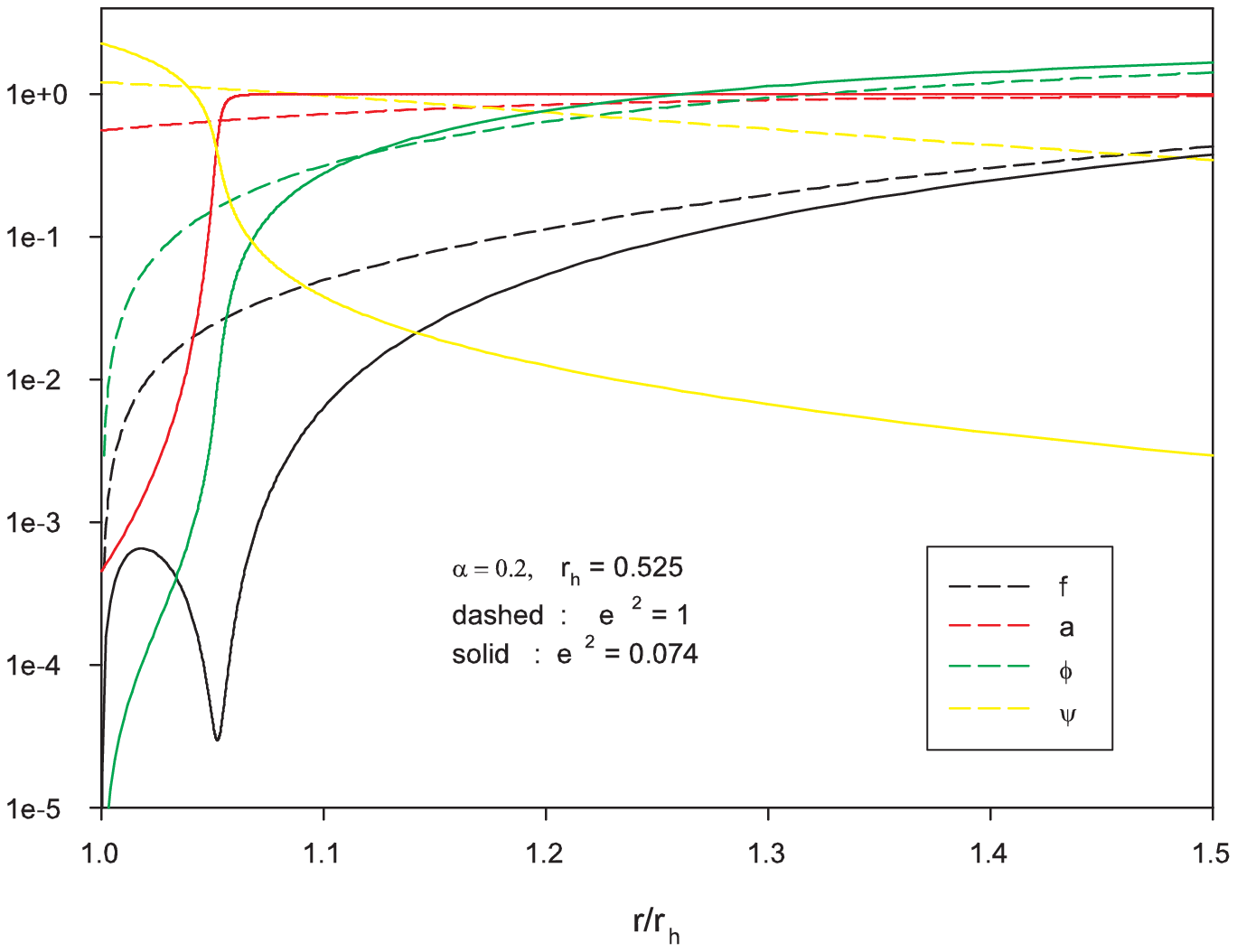}}
\caption{\label{gauss_bonnet_bh_3}
The metric and matter functions of a Gauss-Bonnet black hole solution
for $r_h=0.525$, $\alpha = 0.2$ and two different values of $e^2$ are shown.}
\end{figure}

\subsection{$m^2=0$}
\begin{figure}
\centering
\epsfysize=7cm
\mbox{\epsffile{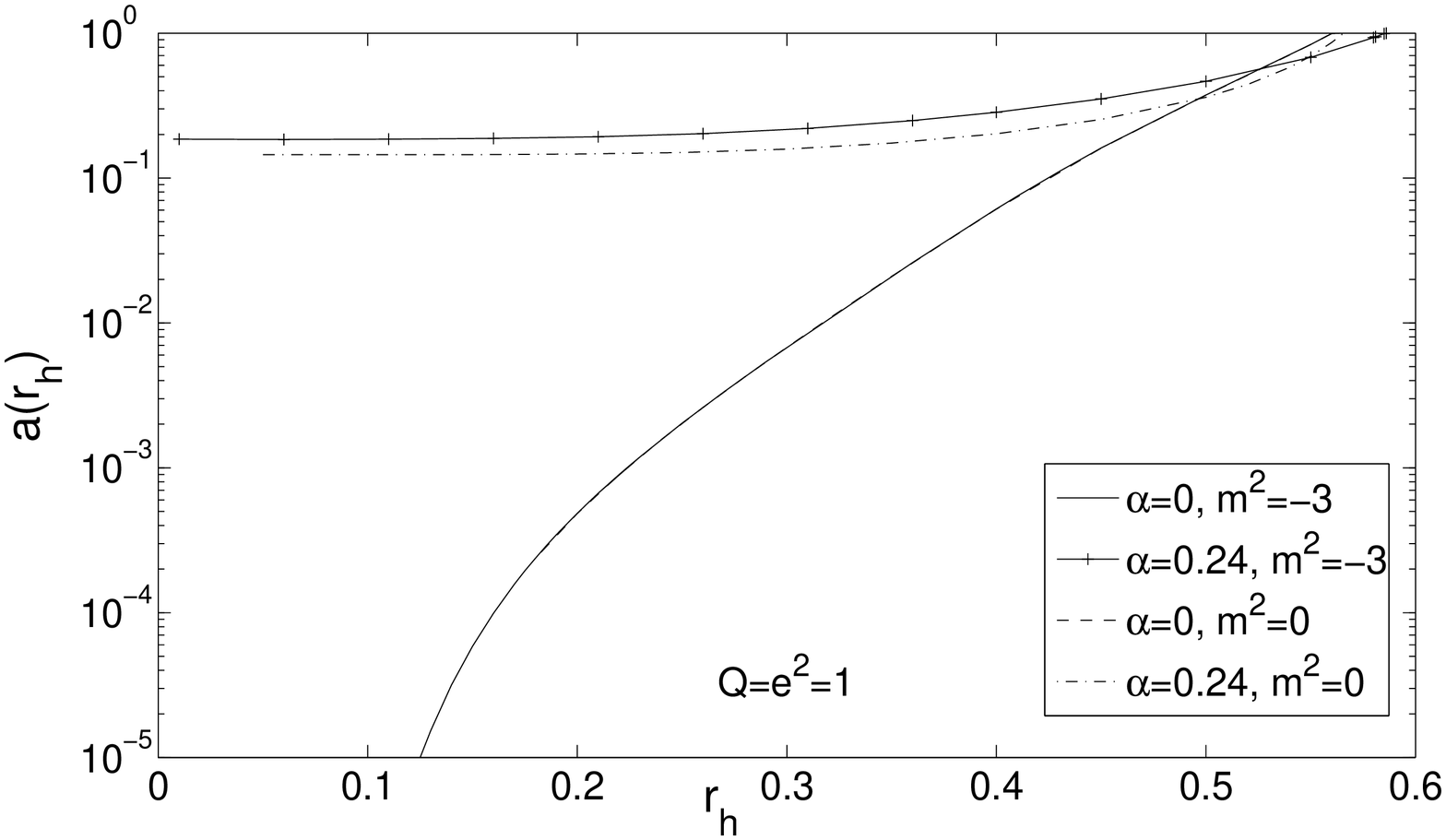}}
\caption{\label{a_xh2}
The value of $a(r_h)$ as function of $r_h$ for a  scalar field with $e^2=1$ and $Q=1$.
We show the curves for $\alpha=0$ and $\alpha=0.24$, respectively as well
as for tachyonic scalar fields with $m^2=-3$ and massless scalar fields with $m^2=0$.
Note that for $\alpha=0$ the curves of the tachyonic and massless case
cannot be distinguished.}
\end{figure}

\subsubsection{Hairy charged black holes for $\alpha=0$}
We have also studied the dependence of our results
on the mass of the scalar field. Our results for $m^2=0$ (together with those for $m^2=-3$) 
are shown
in Fig.\ref{a_xh2}. It is clear from this figure that for $\alpha=0$ the results do not
depend on $m^2$. 

\subsubsection{Hairy charged Gauss-Bonnet black holes}
In Fig.\ref{a_xh2} we also show the dependence of $a(r_h)$ on $r_h$ for $\alpha=0.24$.
Comparing this curve with the corresponding curve for $m^2=-3$, we find that 
$r_{h,max}$ increases with decreasing $m^2$. At some intermediate
$r_h$ the curves for $m^2=0$ and $m^2=-3$ intersect such that for smaller and fixed $r_h$ the
value of $a(r_h)$ increases with decreasing $m^2$. 
In contrast to the $\alpha=0$ case there is hence a dependence on $m^2$.

\section{Summary and Conclusions}

We have studied the instability of static and charged solitons and black holes with respect to the condensation
of a massless and tachyonic scalar field, respectively. We have re-investigated the case of Einstein
gravity and have extended the results to include Gauss-Bonnet gravity terms.
In both cases we have addressed the problem of existence of hairy
      solutions by solving the full system of equations numerically, i.e.
      our solutions are therefore not perturbative solutions.  

For the Einstein case, we find new features as compared to those reported in the perturbative 
limit in \cite{dias2} and propose
the following scenario for the domain of existence of hairy black holes. For fixed $Q$ there exists an $r_{h,ex}$ which corresponds
to the horizon radius of the extremal RNAdS solution. Hence, RNAdS solutions exist only for $r_h > r_{h,ex}$. For 
$r_h=0$ soliton solutions exist only for $e^2 > e_c^2$. We hence find
\begin{itemize}
 \item for fixed (and small) $Q$ and $e^2 > e_c^2$ the black holes tend to the corresponding soliton solution with the same $Q$ when decreasing $r_h$
\item for fixed $Q$ and $e^2 < e_c^2$ the black holes tend to a singular solution when decreasing $r_h$
\item for fixed $Q$ and $r_h > r_{h,ex}$ the black holes tend to the RNAdS solutions with the same charge $Q$ when
decreasing $e^2$
\item for fixed $Q$ and $r_h < r_{h,ex}$ the black holes tend to a configuration that possesses a local minimum of the metric
function $f(r)$ \ .
\end{itemize}

For $\alpha\neq 0$ the scenario is similar except for the fact that the Gauss-Bonnet black holes never tend to the corresponding
soliton solution when decreasing $r_h$ - 
even when the soliton exists for the given values of $Q$ and $e^2$.
It would be interesting to check whether these new features in the
Gauss-Bonnet case  could be recovered within a perturbative
approach comparable to that used in the Einstein case in \cite{dias2}.

Furthermore, none of the limiting black hole solutions can be a regular extremal black hole with
scalar hair. Here we have extended the results of \cite{fiol1} and have shown that
the near-horizon geometry of extremal Gauss-Bonnet black holes does not
support massless or tachyonic scalar hair. 

In extension to \cite{dias2} we observe that for intermediate values of the 
gauge coupling hairy solitons
exist if the charge $Q$ is small enough or large enough and that a ``forbidden band'' of charges
is present in which hairy solitons do not exist.\\
\vspace{1cm}

{\bf Acknowledgments} YB thanks the Belgian FNRS for financial support. We thank the anonymous referee
of our paper for the very constructive and helpful comments.

\section{Appendix: A No-hair theorem for extremal Gauss-Bonnet black holes}
In \cite{fiol1} it was argued that extremal Reissner-Nordstr\"om-AdS black holes
cannot carry scalar hair. Here, we will investigate the case including Gauss-Bonnet corrections.

In the following, we want to discuss the equations in the near-horizon case. It was
shown in \cite{aste} that the near-horizon geometry is AdS$_2\times S^3$. We assume the
metric to have the following form
\begin{equation}
 ds^2=v_1 \left(-\rho^2 d\tau^2 + \frac{1}{\rho^2} d\rho^2\right) + v_2\left(
d\psi^2 + \sin^2\psi\left(d\theta^2 + \sin^2\theta
d\varphi^2\right)\right)  \ ,
\end{equation}
where $v_1$ and $v_2$ are positive constants. 

We find that the equations then read
\begin{equation}
 \label{tt}
\frac{3}{v_2} - \Lambda 
=8\pi G \left( \frac{1}{2} \frac{\phi'^2}{v_1^2} + \frac{\rho^2}{v_1} \psi'^2 + \frac{e^2 \phi^2 \psi^2}{\rho^2 v_1}
+ m^2\psi^2\right)  \ ,
\end{equation}
\begin{equation}
 \label{rr}
-\frac{3}{v_2} + \Lambda 
=8\pi G \left( -\frac{1}{2} \frac{\phi'^2}{v_1^2} + \frac{\rho^2}{v_1} \psi'^2 + \frac{e^2 \phi^2 \psi^2}{\rho^2 v_1}
- m^2\psi^2\right)  \ ,
\end{equation}
\begin{equation}
 \label{thetatheta}
-\frac{1}{v_2} + \Lambda + \frac{1}{v_1}+ \frac{2\alpha}{v_1v_2}
=8\pi G \left( \frac{1}{2} \frac{\phi'^2}{v_1^2} - \frac{\rho^2}{v_1} \psi'^2 
+ \frac{e^2 \phi^2 \psi^2}{\rho^2 v_1}
- m^2\psi^2\right)  \ .
\end{equation}
Note that $H_{\tau\tau}$ and $H_{\rho\rho}$ are vanishing and hence the equations
are the same as for $\alpha=0$ (compare to \cite{fiol1}).

Combination of (\ref{tt}) and (\ref{rr}) yields
\begin{equation}
\label{combi}
0=16\pi G \left(\frac{\rho^2}{v_1} \psi'^2 + \frac{e^2 \phi^2 \psi^2}{\rho^2 v_1}\right) \ .
\end{equation}
We  can hence draw the same conclusion as in the $\alpha=0$ case, namely that
since both terms on the rhs of (\ref{combi}) are positive they must vanish
and hence $\psi'=0$ and $\phi^2 \psi^2=0$ in the near-horizon geometry.
Since $\frac{3}{v_2} - \Lambda > 0$ and $m^2 \leq 0$, the case $\phi\equiv 0$ is ruled
out from (\ref{tt}), hence the only possibility is $\psi\equiv 0$ in the near-horizon
geometry and we conclude that extremal Gauss-Bonnet black holes can {\it not} carry massless
or tachyonic scalar hair. Furthermore (\ref{thetatheta}) is always fulfilled
in this case. To see this note that $v_1$ is equal to the square of the AdS$_2$ radius, i.e.
$v_1=R^2$, which was computed to be given by $R^2=(r_h^2 + 2\alpha)/
(4+ 12r_h^2/L^2)$ \cite{aste,hartmann_brihaye3}. Moreover, $v_2=r_h^2$, where $r_h$ is the 
horizon radius. Inserting these expressions into (\ref{thetatheta}) it is
easy to see that the equation is fulfilled and doesn't put further constraints on the
parameters.


\begin{thebibliography}{30}

\bibitem{ggdual} 
{\it see e.g.} O.~Aharony, S.~S.~Gubser, J.~M.~Maldacena, H.~Ooguri and Y.~Oz,
  Phys.\ Rept.\  {\bf 323} (2000) 183
  [arXiv:hep-th/9905111];
 E.~D'Hoker and D.~Z.~Freedman,
  arXiv:hep-th/0201253;
M. Benna and I. Klebanov, {\it Gauge-string duality and some applications} [arXiv: 0803.1315 [hep-th]].   
\bibitem{adscft} J. Maldacena, Adv. Theo. Math. Phys. {\bf 2} (1998) 231;
Int. J. Theor. Phys. {\bf 38} (1999) 1113 [arXiv:hep-th/9711200].

\bibitem{gubser} S.~S.~Gubser,
  Phys.\ Rev.\  D {\bf 78} (2008) 065034
  [arXiv:0801.2977 [hep-th]].
 \bibitem{hhh} 
 S.~A.~Hartnoll, C.~P.~Herzog and G.~T.~Horowitz, Phys. Rev. Lett. {\bf 101} (2008) 031601 [arXiv:0803.3295 [hep-th]];
  JHEP {\bf 0812} (2008) 015
  [arXiv:0810.1563 [hep-th]].
  
\bibitem{horowitz_roberts} 
  G.~T.~Horowitz and M.~M.~Roberts,
  Phys.\ Rev.\  {\bf D78} (2008) 126008,
  [arXiv:0810.1077 [hep-th]].
\bibitem{reviews} {\it for recent reviews see} C.~P.~Herzog,
  J.\ Phys.\ A  {\bf 42} (2009) 343001;
  S.~A.~Hartnoll,
  Class.\ Quant.\ Grav.\  {\bf 26} (2009) 224002
  [arXiv:0903.3246 [hep-th]];
G. Horowitz, {\it Introduction to holographic superconductors},
arXiv:1002.1722.
 
\bibitem{bf} P.~Breitenlohner and D.~Z.~Freedman,
  Annals Phys.\  {\bf 144} (1982) 249.

\bibitem{Robinson:1959ev}
  I.~Robinson,
  Bull.\ Acad.\ Pol.\ Sci.\ Ser.\ Sci.\ Math.\ Astron.\ Phys.\  {\bf 7} (1959)
351.

\bibitem{Bertotti:1959pf}
  B.~Bertotti,
  Phys.\ Rev.\  {\bf 116} (1959) 1331.




\bibitem{Bardeen:1999px}
  J.~M.~Bardeen, G.~T.~Horowitz,
  Phys.\ Rev.\   {\bf D60 } (1999)  104030,
  [arXiv:hep-th/9905099].





\bibitem{Sen:2005wa}
  A.~Sen,
  JHEP {\bf 0509 } (2005)  038,
  [arXiv:hep-th/0506177].


\bibitem{sen2}
  A.~Sen,
  JHEP {\bf 0811 } (2008)  075,
  [arXiv:0805.0095 [hep-th]].

\bibitem{dias_silva} 
  O.~J.~C.~Dias, P.~J.~Silva,
  Phys.\ Rev.\  {\bf D77 } (2008)  084011,
  [arXiv:0704.1405 [hep-th]].

\bibitem{Dias:2010ma}
  O.~J.~C.~Dias, R.~Monteiro, H.~S.~Reall, J.~E.~Santos,
  JHEP {\bf 1011 } (2010)  036,
  [arXiv:1007.3745 [hep-th]].


\bibitem{zwiebach}  B.~Zwiebach,
  Phys.\ Lett.\   {\bf B156 } (1985)  315;  R.~I.~Nepomechie,
  Phys.\ Rev.\   {\bf D32 } (1985)  3201.


\bibitem{deser}
  D.~G.~Boulware, S.~Deser,
  Phys.\ Rev.\ Lett.\  {\bf 55 } (1985)  2656.

\bibitem{Wheeler:1985nh}
  J.~T.~Wheeler,
  Nucl.\ Phys.\  {\bf B268 } (1986)  737.
  


\bibitem{wiltshire}
  D.~L.~Wiltshire,
  Phys.\ Rev.\  {\bf D38 } (1988)  2445.
  
\bibitem{Cai:2001dz}
  R.~-G.~Cai,
  Phys.\ Rev.\   {\bf D65} (2002)  084014,
  [arXiv:hep-th/0109133].

\bibitem{Cvetic:2001bk}
  M.~Cvetic, S.~'i.~Nojiri, S.~D.~Odintsov,
  Nucl.\ Phys.\ {\bf B628 } (2002)  295,
  [arXiv:hep-th/0112045].

\bibitem{Cho:2002hq}
  Y.~M.~Cho, I.~P.~Neupane,
  Phys.\ Rev.\  {\bf D66 } (2002)  024044,
  [arXiv:hep-th/0202140].

\bibitem{Neupane:2002bf}
  I.~P.~Neupane,
  Phys.\ Rev.\  {\bf D67} (2003)  061501,
  [arXiv:hep-th/0212092].

\bibitem{Cai:2003gr}
  R.~-G.~Cai, Q.~Guo,
  Phys.\ Rev.\  {\bf D69 } (2004)  104025,
  [arXiv:hep-th/0311020].

\bibitem{Neupane:2003vz}
  I.~P.~Neupane,
  Phys.\ Rev.\  {\bf D69 } (2004)  084011,
  [arXiv:hep-th/0302132].

\bibitem{Clunan:2004tb}
  T.~Clunan, S.~F.~Ross, D.~J.~Smith,
  Class.\ Quant.\ Grav.\  {\bf 21 } (2004)  3447,
  [arXiv:gr-qc/0402044].

\bibitem{hartmann_brihaye3}  Y.~Brihaye and B.~Hartmann,
  Phys.\ Rev.\ D {\bf 84} (2011) 084008
  [arXiv:1107.3384 [gr-qc]].

\bibitem{CMW} N.~D.~Mermin, H.~Wagner, 
Phys. Rev. Lett. {\bf 17} (1966) 1133;
S.~Coleman,
Commun. Math. Phys. {\bf 31} (1973) 259.

\bibitem{Gregory:2009fj}
  R.~Gregory, S.~Kanno, J.~Soda,
  JHEP {\bf 0910 } (2009)  010,
  [arXiv:0907.3203 [hep-th]].

\bibitem{Brihaye:2010mr}
  Y.~Brihaye, B.~Hartmann,
  Phys.\ Rev.\  {\bf D81 } (2010)  126008,
  [arXiv:1003.5130 [hep-th]].


\bibitem{Barclay:2010up}
  L.~Barclay, R.~Gregory, S.~Kanno, P.~Sutcliffe,
  JHEP {\bf 1012 } (2010)  029,
  [arXiv:1009.1991 [hep-th]].

\bibitem{Siani:2010uw}
  M.~Siani,
  JHEP {\bf 1012 } (2010)  035.
  [arXiv:1010.0700 [hep-th]].

\bibitem{dias2} O. Dias, P. Figueras, S. Minwalla, P. Mitra, R. Monteiro, J. Santos, {\it Hairy black holes
and solitons in global AdS$_5$}, arXiv: 1112.4447 [hep-th].

\bibitem{basu}
 P.~Basu, J.~He, A.~Mukherjee and H.~H.~Shieh,
  JHEP {\bf 0911}, 070 (2009)
  [arXiv:0810.3970 [hep-th]].

\bibitem{fiol1}
J.~Fernandez-Gracia and B.~Fiol,
  JHEP {\bf 0911} (2009) 054
  [arXiv:0906.2353 [hep-th]].
\bibitem{brihaye_hartmannNEW} 
 Y.~Brihaye and B.~Hartmann,
{\it A Scalar field instability of rotating and charged black holes in 
(4+1)-dimensional Anti-de Sitter space-time}, JHEP (2012), in press,
  arXiv:1112.6315 [hep-th].

\bibitem{menagerie} S.~A.~Gentle, M.~Rangamani and B.~Withers,
 {\it A Soliton Menagerie in AdS},
  arXiv:1112.3979 [hep-th].

\bibitem{colsys} U. Ascher, J. Christiansen and R. D. Russell, Math. Comput. {\bf 33}
(1979), 659; ACM Trans. Math. Softw. {\bf 7} (1981), 209.

\bibitem{aste} 
  D.~Astefanesei, N.~Banerjee, S.~Dutta,
  JHEP {\bf 0811 } (2008)  070,
  [arXiv:0806.1334 [hep-th]].


\end{thebibliography}
\end{document}